\documentclass[11pt,a4paper,twoside]{article}        
\makeatletter
\usepackage{amsbsy}
\usepackage{amsfonts}
\usepackage[centertags]{amsmath} 
\usepackage{amsopn}
\usepackage{amssymb}
\usepackage{amstext}
\usepackage{graphicx}
\usepackage{array}
\usepackage{boxedminipage}
\usepackage{calc}
\usepackage{cite}
\usepackage[dvips]{color}
\usepackage{curves}
\usepackage{dsfont}
\usepackage{epsfig,epic}
\usepackage{eepic}
\usepackage{fancyhdr}
\usepackage{float}
\usepackage{latexsym}   
                        
\usepackage{rotating}    
\usepackage{setspace}
\usepackage{subfigure}
\usepackage{syntonly}   
\usepackage{theorem} 
\usepackage{wrapfig}
\usepackage[latin1]{inputenc}
\usepackage[T1]{fontenc}
\usepackage[english]{babel}
\newlength{\rilegatura}
\newlength{\simmetricspaceh}
\newlength{\fotogram}
\newlength{\sepfoto}
\newlength{\larggraph}
\newlength{\altgraph}
\DeclareMathOperator*{\Tr}{Tr}

\DeclareMathOperator*{\slim}{s--lim}
\newcommand{\Lspace}[2]{L^{#1}_{\mu}\pt{#2}} 
\newcommand{\qed}{\rule[0pt]{7pt}{7pt}} 
\newcommand{\Cspace}[2]{{\cal C}^{#1}\pt{#2}}  
  
\def\IN{\mathds{N}}  
\def\IZ{\mathds{Z}}  
\def\IR{\mathds{R}}  
  
\def\pum{{\scriptstyle \frac{1}{2}}}  
\def\IT{{\mathds{T}}^2}

\def\Ac{{\cal C}^{0}\pt{\IT}}      
\def\Al{L^{\infty}_{\mu}\pt{\IT}}  
\def\om{\omega_\mu}  
\def\tn{\tau_{\cal N}}      
\def\car{{\cal X}}      
\def\nh{{\cal N}}      
\def\ZNZD{{(\IZ/ N \IZ)}^2}	
	        
\def\tripAS{\big(L^{\infty}_{\mu}\pt{\IT},\omega_\mu,\Theta)}
\def\tripASa{\big(L^{\infty}_{\mu}\pt{\IT},\omega_\mu,\Theta_\alpha\big)}
\def\tripQS{\big({\cal D}_\nh,\tn,\Theta_{\nh}\big)}
\def\tripQSa{\big({\cal D}_\nh,\tn,{\Theta}_{\nh,\alpha}\big)}
\def\coleq{\raisebox{0.0815\height}{$\colon$}\!\!\!\!=}         
							        
\def\eqcol{=\!\!\!\!\!\!\:\ \raisebox{0.0815\height}{$\colon$}}

\newcommand{\ZNZ}[1]{\IZ/#1\IZ}

\newcommand{\newatop}[2]{\genfrac{}{}{0pt}{}{#1}{#2}}   
							 
\newcommand{\pt}[1]{\left( #1 \right)}                   
\newcommand{\pq}[1]{\left[ #1 \right]}                   
\newcommand{\pg}[1]{\left\{ #1 \right\}}                 
\newcommand{\floor}[1]{\left\lfloor #1 \right\rfloor}             
                
\newcommand{\abs}[1]{\left|\left. #1 \right.\right|}              
\newcommand{\bk}[1]{\left\langle #1 \right\rangle}                
  
\newcommand{\bkkk}[3]{\left.\left\langle#1\left|#2\right|
                                          #3\right\rangle\right.} 
\newcommand{\ket}[1]{\left|\left. #1 \right.\right\rangle}        
\newcommand{\bra}[1]{\left\langle\left. #1 \right.\right|}        
\newcommand{\norm}[2]{{\left\Arrowvert #1 \right\Arrowvert}_{#2}} 
\newcommand{\enfasi}[1]{{\it #1}}          
\newcommand{\bs}[1]{\boldsymbol{#1}}

\newcommand{\ud}{\mathrm{d}}

\renewcommand{\Re}{\mathrm{Re}}
\renewcommand{\geq}{\geqslant}       
\renewcommand{\leq}{\leqslant}

\newcommand{\Ir}{\mathds{Z}}

\newcommand{\idty}{\mathds{1}}

\newcommand{\coh}[2]{\ensuremath{|C_{#1}(#2) \rangle}}
\newcommand{\lcoh}[2]{\ensuremath{\langle C_{#1}(#2)|}}
\newcommand{\<}{\langle}
\renewcommand{\>}{\rangle}

\renewcommand{\c}[1]{\mathcal{#1}}

\newcommand{\co}[1]{\textsf{#1}}
{\theoremstyle{plain} 
\newtheorem{TT}{Theorem}
}
{\theoremstyle{plain} \theorembodyfont{\rmfamily}
} 
{\theoremstyle{plain} \theorembodyfont{\rmfamily}
} 
{\theoremstyle{break} \theorembodyfont{\rmfamily}

} 
{\theoremstyle{break} \theorembodyfont{\rmfamily}
} 
{\theoremstyle{break} \theorembodyfont{\rmfamily}
} 
{\theoremstyle{plain} \theorembodyfont{\rmfamily}
} 
{\theoremstyle{plain} \theorembodyfont{\rmfamily}
} 
{\theoremstyle{break} \theorembodyfont{\rmfamily}
} 
{\theoremstyle{break} \theorembodyfont{\rmfamily}
\newtheorem{PRS}{Properties}[section]} 
{\theoremstyle{break} \theorembodyfont{\rmfamily}
\newtheorem{LLL}{Lemma}[section]} 
{\theoremstyle{break} \theorembodyfont{\rmfamily}
\newtheorem{NNS}{Remarks}[section]
\newtheorem{NNN}[NNS]{Remark}} 
{\theoremstyle{break} \theorembodyfont{\rmfamily}
} 
{\theoremstyle{break} \theorembodyfont{\rmfamily}
\newtheorem{DDD}{Definition}[section]
\newtheorem{DDS}[DDD]{Definitions}} 
{\theoremstyle{break} \theorembodyfont{\rmfamily}
\newtheorem{PPP}{Proposition}[section]
}

\newenvironment{Ventry}[1]
	{\begin{list}{}{
		\settowidth{\labelwidth}{#1}
		\setlength{\leftmargin}{\labelwidth}}}
	{\end{list}}
\renewcommand\section{\@startsection{section}{1}{\z@}
                            {-3.5ex \@plus -1ex \@minus -.2ex}
                            {2.3ex \@plus.2ex}
                            {\normalfont\Large\bfseries\boldmath}}
\renewcommand\subsection{\@startsection{subsection}{2}{\z@}
                            {-3.25ex\@plus -1ex \@minus -.2ex}
                            {1.5ex \@plus .2ex}
                            {\normalfont\large\bfseries\boldmath}}
\renewcommand\subsubsection{\@startsection{subsubsection}{3}{\z@}
                            {-3.25ex\@plus -1ex \@minus -.2ex}
                            {1.5ex \@plus .2ex}
                            {\normalfont\normalsize\bfseries\boldmath}} 
\renewcommand\paragraph{\@startsection{paragraph}{4}{\z@}
                            {3.25ex \@plus1ex \@minus.2ex}
                            {-1em}
                            {\normalfont\normalsize\bfseries\boldmath}}
\renewcommand\subparagraph{\@startsection{subparagraph}{5}{\parindent}
                             {3.25ex \@plus1ex \@minus.2ex}
                             {-1em}
                             {\normalfont\normalsize\bfseries\boldmath}}  
\renewcommand{\@seccntformat}[1]{\csname the#1\endcsname.\hspace{1em}}

\definecolor{cyan}{cmyk}{1,0.1,0,0.2}

\newcommand{\ccro}{\color{red}}

\newcommand{\ccbl}{\color{blue}}

\newcommand{\ccvi}{\color{magenta}}

\newcommand{\nncc}{\normalcolor}
\makeatother
\title{\bf Continuous Limit of Discrete Sawtooth Maps and its Algebraic
Framework} 
\author{FABIO BENATTI\\
Dipartimento di Fisica Teorica\\
Universit\`a di Trieste\\
and Istituto Nazionale di Fisica Nucleare, Sezione di Trieste,\\
Strada Costiera 11, 34014 Trieste, Italy\\
fabio.benatti@ts.infn.it\\[3ex]
\and
VALERIO CAPPELLINI\\
Dipartimento di Fisica Teorica\\
Universit\`a di Trieste\\
Strada Costiera 11, 34014 Trieste, Italy\\
valerio.cappellini@ts.infn.it}
\begin{document}    
\singlespacing
\maketitle
\begin{abstract}
\noindent We study the presence of a logarithmic time scale in
discrete approximations of Sawtooth Maps on the 2--torus. The techniques
used are suggested by quantum mechanical similarities, and are based
on a particular class of states on the torus, that fulfill
dynamical localization properties typical of quantum Coherent States.\\[2ex]
Mathematics Subject Classification $2000$: 37D20; 81Q20, 81R30, 46L99
\end{abstract}
\pagestyle{fancy}
\lhead[\fancyplain{}{\footnotesize \thepage \protect\hspace{5mm} \it
F.~Benatti and V.~Cappellini }]
      {\fancyplain{}{}}
\rhead[\fancyplain{}{}]
      {\fancyplain{}{\footnotesize \it Continuous Limit of Discrete
Sawtooth Maps and its Algebraic Framework \hspace{5mm}\thepage}}
\chead{}\lfoot{}\cfoot{}\rfoot{}
\tableofcontents
\onehalfspacing
\section{Introduction}
\label{intro}
Under the
term of Quantum Chaos goes a rich phenomenology of
behaviours~\cite{Zas85:1,Gia89:1,Cas95:1} 
proper to quantum systems whose classical limit presents typical
chaotic features as
positive Lyapunov exponents (hyperbolic
regime)~\cite{Dev89:1,Wig90:1,Kat99:1}.  

The footprints of classical chaos are usually studied
semi--classically when a suitable ``$\hbar$''--like quantization parameter
goes to zero; one then examines the differences between quantum and 
classical behaviours.
In the hyperbolic case, quantum chaos reveals itself through
the presence of a time--scale, over which quantum and classical
motions mimic each other, that increases as
$-\log\hbox{``$\hbar$''}$~\cite{Zas85:1,Gia89:1,Cas95:1,Bou04:1,Fnj04:1,Sch04:1}. 
This peculiar logarithmic time scale has to be compared with the
scaling $\hbox{``$\hbar$''}^{-\alpha},\ \alpha>0$, which is proper of
quantum systems with regular classical limit~\cite{Zas85:1}.

Heuristical explanations of the logarithmic time-scale already
indicate that the phenomenon is not exclusive of quantum systems, and
thus of non--commutativity,
but that it should also be present when the classical dynamics is
looked at as the continuous limit of a family of discrete classical
systems.~\cite{Cap04:1}.   
 
Intrinsically discrete systems~\cite{Cri93:1} and discretized
classical continuous 
systems~\cite{Cri94:1,Fal03:1,Ben04:1} have recently been objects of
numerical analysis concerning 
the entropy 
production and the presence of a
logarithmic time scale, whereas the ergodic properties of discretized
discontinuous maps have been addressed in~\cite{Mar05:1}.

In the following, we shall rigorously show this fact to be true for
\co{S}awtooth \co{M}aps on the $2$-dimensional
torus~\cite{Che92:1,Vai92:1,Per87:1}: this will be done
by forcing them to move on a square
lattice and by retrieving the continuous dynamics 
when the lattice spacing goes to zero. 
Because of the analogies between quantization and discretization, we
will make use of technologies strictly resembling the so-called
{\it Anti--Wick} quantization~\cite{Bon03:1}.

We shall prove that a time--scale logarithmic in the lattice-spacing 
appears; in comparison to previous
results obtained studying numerically the entropy
production~\cite{Ben04:1}, a rigorous 
continuous limit is established that succeeds in controlling the
discontinuities of \co{S}awtooth \co{M}aps.
Despite their classical nature, the entropy
previously investigated was quantum mechanical; somewhat
analogously, in this article, \co{S}awtooth \co{M}aps will be studied by
means of states, which play a role similar to quantum Coherent States, whose choice is naturally provided by the lattice
structure of discretized \co{S}awtooth \co{M}aps.
They will be shown to satisfy a {\it dynamical localization property}
that makes them remain localized around the trajectories of the
continuous dynamics, but only on a logarithmic time scale.
\section{Classical Dynamical Systems}
\label{CDS}
Classical dynamics is usually described by means of a
measure space ${\cal X}$, the phase--space, endowed with the Borel
$\sigma$--algebra and a normalized 
measure $\mu$, $\mu({\cal X})=1$.
The ``volumes'' 
\begin{equation*}
\mu(E)=\int_E\mu(\ud \bs{x})
\end{equation*}
of measurable subsets 
$E\subseteq{\cal X}$ represent the probabilities that a
phase--point $\bs{x}\in{\cal X}$ belong to them:
the measure $\mu$ defines the statistical properties of the system and
represents a possible state, which is taken to be an equilibrium
state with respect to the given dynamics.

In such a scheme, a reversible discrete time dynamics amounts to an
invertible measurable 
map $S:{\cal X}\mapsto{\cal X}$ such that $\mu\circ S=\mu$ and to its
iterates $\{S^k \mid k\in\Ir\}$:
phase--trajectories passing through $\bs{x}\in{\cal X}$ at time $0$
are then sequences ${\pg{S^k\,\bs{x}}}_{k\in\IZ}$~\cite{Kat99:1}.

Classical dynamical systems are thus conveniently described by
triplets $({\cal X},\mu,S)$; in the present work, we shall focus upon
the following choices: 
\begin{itemize}
\item[$\cal X$:] the $2$--dimensional
 torus $\IT={\IR}^2/{\IZ}^2 \-=\left\{\bs{x}=(x_1,x_2)\in\IR^2\
 \pmod{1} \right\}$;
\item[$\mu$:] the Lebesgue measure, $\mu(\ud\bs{x})=\ud x_1\,\ud x_2$,
on $\IT$; 
\item[$S$:] an invertible measurable transformations on $\IT$ that
preserves the Lebesgue measure.
\end{itemize}

It is convenient to associate an algebraic triple $({\cal
M},\omega,\Theta)$ to the measure--theoretic triple $(\IT,\mu,S)$,
consisting of 
\begin{itemize}
	\item[${\cal M}$:] the (Abelian) Von~Neumann *-algebra
	$\Lspace{\infty}{\IT}$ of essentially bounded functions on
	$\IT$~\cite{Bra79:1,Hew69:1}.
	\item[$\omega_{\mu}$:] the state (expectation) on ${\cal M}$,
	given by 
	\begin{equation}
	\om:\Lspace{\infty}{\IT}\ni
	f\longmapsto\omega_\mu(f)\coleq\int_{\IT} \mu(\ud\bs{x})\
	f(x)\in \IR^{+}\ \cdot
	\label{omegamu}
	\end{equation}
	\item[$\Theta$:] the
        automorphism
        of ${\cal M}$ such that $\Theta\pt{f}=f\circ S$,
        $\omega\circ\Theta=\omega$.  
\end{itemize}
In the following, we shall consider a discretized version of
$(\IT,\mu,S)$ which arises by forcing the continuous classical system
to live on a square lattice $L_N\subseteq\IT$ of spacing
$\frac{1}{N}$:
\begin{equation}
L_N \coleq \pg{\frac{\bs{p}}{N} \ \Big|\  \bs{p}\in {\pt{\IZ / N
\IZ}}^2}\ ,
\label{llnn}
\end{equation}
where $\pt{\IZ / N \IZ}$ denotes the residual
class$\pmod{N}$, that is $0 \leqslant p_i \leqslant N-1$.

Taking the $N^2$ points as labels of the elements 
	${\pg{\ket{\bs{\ell}}}}_{\bs{\ell}\in   
	{\pt{\IZ / N \IZ}}^2}$ of an orthonormal basis (o.n.b.) of the ${\cal
	N}$ dimensional  
	Hilbert space ${\cal H}_{\nh}$, ${\cal
	N}\coleq N^2$, we will consider
	discrete algebraic triples $\tripQS$, consisting of
\begin{itemize}
\item[${\cal D}_{\nh}$:] an $\nh \times \nh$ matrix algebra diagonal
in the orthonormal basis introduced above;
\item[$\tn$:] the uniform state (expectation) on $\cal D_{\nh}$ defined by 
\begin{equation}
\tn:{\cal D}_\nh\ni
D\longmapsto\tn(D)\coleq\frac{1}{\nh}\Tr\pt{D}\in \IR^{+}\ ;
\label{tauenne}
\end{equation}
\item[$\Theta_{\nh}$:] an automorphism of $\cal D_{\nh}$ suitably
reproducing 
$\Theta$ when $N\longrightarrow\infty$ (see Section~\ref{AdoT}).
\end{itemize}
\begin{quote}
\begin{NNN}{}\ \\[-5.5ex]
\begin{Ventry}{}
\item As it will become evident in the following, up to a certain
extent, discretization resembles quantization; in the latter case,
instead of ${\cal D}_\nh$, one deals with
non--commutative matrix algebras, the typical instance being the
finite dimensional quantization of the Arnold Cat Map~\cite{Ber80:1,Deg93:1}.
\end{Ventry}
\end{NNN}
\end{quote}
\section{Discretization of phase--space}
\label{dops}
\vspace{6mm}
As sketched in the previous Remark, we proceed now
to setup a discretization procedure close to the so--called
\co{A}nti--\co{W}ick quantization~\cite{Bon03:1}. 

Given the classical algebraic triple $\tripAS$, the aim of a 
discretization--dediscretization procedure (specifically an
$\nh$--dimensional discretization) is twofold:
\begin{itemize}
\item finding a pair of *-morphisms, ${\cal J}_{\nh,\infty}$ mapping
$\Lspace{\infty}{\IT}$ into the
abelian finite dimensional algebra ${\cal D}_\nh$ and ${\cal
J}_{\infty,\nh}$ mapping backward ${\cal D}_\nh$ into
$\Lspace{\infty}{\IT}$; 
\item providing an automorphism $\Theta_{\nh}$, the discrete dynamics,
acting on ${\cal D}_\nh$ 
such that it approximates the continuous one, $\Theta$, on
$\Lspace{\infty}{\IT}$ as follows
\begin{equation}
{\cal J}_{\infty,\nh}^{\phantom{j}}\circ\Theta_{\nh}^j\circ {\cal
J}_{\nh,\infty}^{\phantom{j}}\xrightarrow[N\to\infty]{}
 \Theta_{\phantom{\nh}}^j\ \cdot 
\label{commJT}
\end{equation}
\end{itemize}
The latter requirement can be seen as a modification of the so called
Egorov's property (see~\cite{Mar99:1}).  
Intuitively, a discrete description of the measure--theoretic triple 
 $\big(\IT,\mu,S\big)$ becomes finer when we increase $N$, the number
of points per linear dimension on the grid $L_N$ in~\eqref{llnn}:
this corresponds to enlarging the dimension of the Hilbert space $\cal
H_\nh$ associate to the corresponding algebraic triple $\tripQS$.
In this sense, the lattice
 spacing $a\coleq\frac{1}{N}$ of the grid $L_N$ is a natural
``discretization parameter'' playing an analogous role to the
quantization parameter $\hbar$. 

The difficulty is to find convenient *-morphisms  ${\cal
J}_{\nh,\infty}$ and ${\cal J}_{\infty,\nh}$ that set up a 
rigorous asymptotic (in $N$) correspondence, of 
functions on $\Lspace{\infty}{\IT}$ and matrices in ${\cal D}_\nh$
and, above all, between the discrete dynamics $\Theta_{\nh}$ and the
continuous one $\Theta$.

Due to the similarities with quantization, we shall consider 
a discretization
procedure based on states that we shall refer to as \co{L}attice
\co{S}tates (\co{LS} for short) which mimic the use of \co{C}oherent
\co{S}tates in the study of the semi--classical limit.
In the next section we will give a
suitable definitions of \co{LS} belonging to the Hilbert space $\cal
H_\nh$, that we shall use to discretize $\tripAS$. 
\subsection{\co{L}attice \co{S}tates on $\IT$}
\label{PSCS}
\vspace{3mm}  
In analogy with the the properties of quantum
\co{C}oherent \co{S}tates,  
we shall look for a class $\{\vert C_\nh(\bs{x})\rangle \mid
\bs{x}\in\IT\}\in {\cal H}_\nh$ of vectors, 
 indexed by points
 $\bs{x}\in\IT$, satisfying the following conditions which are
borrowed from analogous quantum ones~\cite{Ben03:1}:
\begin{quote}
\begin{PRS}{}\ \\[-6.5ex]
\label{coh}
 \begin{enumerate}
 \item
  \co{Measurability}: $\bs{x} \mapsto \vert C_\nh(\bs{x})\rangle$ is measurable on
 $\IT$;\\[-1ex]
 \item
  \co{Normalization}: $\|C_\nh(\bs{x})\|^2 = 1$, $\bs{x}\in\IT$;\\[-1ex]
 \item
  \co{Completeness}: $\displaystyle \nh \int_{\IT}\mu(\ud\bs{x})\, \coh{\nh}{\bs{x}}
  \lcoh{\nh}{\bs{x}} = \idty$;\\[-1ex]
 \item
  \co{Localization}: given $\varepsilon>0$ and $d_0>0$, there exists 
  $N_0(\epsilon,d_0)$ such that for $N\ge N_0(\epsilon,d_0)$ and
 $d_{\IT}(\bs{x},\bs{y})\ge d_0$ one has
  \begin{equation*} 
   \nh \;|\< C_\nh(\bs{x}), C_\nh(\bs{y}) \>|^2 \le \varepsilon.
  \end{equation*}
 \end{enumerate}
\end{PRS}
\end{quote}
\noindent
The symbol $d_{\IT}(\bs{x},\bs{y})$ used in the \co{localization} property 
stands for the length of the shorter segment
connecting the two points $\bs{x},\bs{y}\in \IT$, namely\\[-2ex]
\begin{quote}
\begin{DDD}{}\ 
\label{Gnbar_m2}
\\[-1ex]
\label{dont}
We shall denote by $\displaystyle
d_{\IT}\pt{\bs{x},\bs{y}}  \coleq
\min_{\bs{n}\in{\IZ}^2} \norm{\bs{x}-\bs{y}+\bs{n}}{{\IR}^2}$
the distance on ${\IT}$.
\end{DDD}
\end{quote}

We shall now construct a family of $\vert C_\nh(\bs{x})\rangle$. Let 
$\floor{\cdot}$ denote the integer part of a real number, namely
$x-1 <\floor{x}\leq x$ is the largest integer smaller than $x$;
further, let $\bk{\cdot}$ denote
the fractional parts, that is $\bk{x}\coleq x - \floor{x}$. Thus we
will write
\begin{equation}
\IT\ni\bs{x} =\pt{\frac{\floor{N
x_1}}{N},\frac{\floor{N x_2}}{N}} + 
\pt{\frac{\bk{N x_1}}{N},\frac{\bk{N x_2}}{N}}\ ,
\notag
\end{equation}
or, more compactly, $\displaystyle \bs{x} =\frac{\floor{N
\bs{x}}}{N} + \frac{\bk{N \bs{x}}}{N}$.
We proceed by associating to points of $\IT$ specific lattice points.
\begin{quote}
\begin{DDD}[\co{L}attice \co{S}tates]{}\ \\
\label{xxxnnn}
Given $\bs{x}\in\IT$, we shall denote by $\hat{\bs{x}}_N$ the element of
$\ZNZD$ given by
\begin{equation}
\hat{\bs{x}}_N=\pt{\hat{x}_{N,1},\hat{x}_{N,2}} \coleq\Big(\floor{N
x_1 + \pum}\,,\,\floor{N x_2 
+ \pum}\Big) \ ,
\label{loc_c32}
\end{equation}
and call \co{L}attice \co{S}tates on $\IT$ the vectors
$\coh{\nh}{\bs{x}}$ defined by
\begin{equation}
\IT\ni\bs{x} \mapsto  \vert C_{\nh}(\bs{x})\rangle\coleq
\ket{\hat{\bs{x}}_N}
\in {\cal H}_\nh\ \cdot
\label{CSforL1}
\end{equation}
\end{DDD}\ \\[-9.5ex]
\begin{NNN}{}\ \\[-5.5ex]\label{distCS}
\begin{Ventry}{}
\item The family of states $|C_{\nh}(\bs{x})\rangle$ is constructed
by choosing, for each $\bs{x}\in\IT$,
that element of the basis of ${\cal H}_\nh$ which is labeled by
the closest element of $L_N$ to
$\bs{x}$. 
\end{Ventry}
\end{NNN}
\end{quote}\ \\[-10.5ex]
\begin{figure}[H]
\begin{center}
\begin{picture}(52,80)(-2,-12) 

\thicklines
\thinlines
\matrixput(0,0)(12,0){6}(0,12){6}{\circle{2}}
\matrixput(1,0)(12,0){5}(0,12){6}{\line(1,0){10}}
\matrixput(0,1)(12,0){6}(0,12){5}{\line(0,1){10}}
\thicklines
\put(0,0){\includegraphics[width=6\unitlength,height=6\unitlength]{./bb1.epsi}}
\put(0,0){\circle*{2}}
\dottedline{2}(6,0)(6,6)
\dottedline{2}(0,6)(6,6)
\put(0,54){\includegraphics[width=6\unitlength,height=6\unitlength]{./bb1.epsi}}
\put(0,60){\circle*{2}}
\put(0,54){\line(1,0){6}}
\dottedline{2}(6,54)(6,60)
\put(54,0){\includegraphics[width=6\unitlength,height=6\unitlength]{./bb1.epsi}}
\put(60,0){\circle*{2}}
\put(54,0){\line(0,1){6}}
\dottedline{2}(54,6)(60,6)
\put(54,54){\includegraphics[width=6\unitlength,height=6\unitlength]{./bb1.epsi}}
\put(60,60){\circle*{2}}
\put(54,54){\line(0,1){6}}
\put(54,54){\line(1,0){6}}
\put(18,54){\includegraphics[width=12\unitlength,height=6\unitlength]{./bb2.epsi}}
\put(24,60){\circle*{2}}
\put(18,54){\line(0,1){6}}
\put(18,54){\line(1,0){12}}
\dottedline{2}(30,54)(30,60)
\put(18,0){\includegraphics[width=12\unitlength,height=6\unitlength]{./bb2.epsi}}
\put(24,0){\circle*{2}}
\put(18,0){\line(0,1){6}}
\dottedline{2}(30,0)(30,6)
\dottedline{2}(18,6)(30,6)
\put(0,18){\includegraphics[width=6\unitlength,height=12\unitlength]{./bb3.epsi}}
\put(0,24){\circle*{2}}
\put(0,18){\line(1,0){6}}
\dottedline{2}(6,18)(6,30)
\dottedline{2}(0,30)(6,30)
\put(54,18){\includegraphics[width=6\unitlength,height=12\unitlength]{./bb3.epsi}}
\put(60,24){\circle*{2}}
\put(54,18){\line(1,0){6}}
\put(54,18){\line(0,1){12}}
\dottedline{2}(54,30)(60,30)
\put(30,30){\includegraphics[width=12\unitlength,height=12\unitlength]{./bb4.epsi}}
\put(36,36){\circle*{2}}
\dottedline{2}(42,30)(42,42)
\dottedline{2}(30,42)(42,42)
\put(30,30){\line(0,1){12}}
\put(30,30){\line(1,0){12}}
\put(-4,-12){\makebox(8,8)[cc]{$0$}}
\put(8,-12){\makebox(8,8)[cc]{$\frac{1}{5}$}}
\put(20,-12){\makebox(8,8)[cc]{$\frac{2}{5}$}}
\put(32,-12){\makebox(8,8)[cc]{$\frac{3}{5}$}}
\put(44,-12){\makebox(8,8)[cc]{$\frac{4}{5}$}}
\put(56,-12){\makebox(8,8)[cc]{$1$}}
\put(-12,-4){\makebox(8,8)[cc]{$0$}}
\put(-12,8){\makebox(8,8)[cc]{$\frac{1}{5}$}}
\put(-12,20){\makebox(8,8)[cc]{$\frac{2}{5}$}}
\put(-12,32){\makebox(8,8)[cc]{$\frac{3}{5}$}}
\put(-12,44){\makebox(8,8)[cc]{$\frac{4}{5}$}}
\put(-12,56){\makebox(8,8)[cc]{$1$}}
\end{picture}
\caption{The above picture represents a square lattice ($L_5$) of spacing 
$\frac{1}{5}$ by circles and connecting lines. All
points in the blue square
$I_{\pt{\frac{3}{5},\frac{3}{5}}}
\protect\coleq\left[\frac{5}{10},\frac{7}{10}\right)\times
\left[\frac{5}{10},\frac{7}{10}\right)\subset\IT$ are
associated with the grid point
$\pt{\frac{3}{5},\frac{3}{5}}$ (black dot).
Thus, for all $\bs{x}\in I_{\pt{\frac{3}{5},\frac{3}{5}}}$,
it turns out that
$|C_{\nh}(\bs{x})\rangle = |\pt{3,3}\rangle\in{\cal H}_\nh$.} 
\label{lattice}
\end{center}
\end{figure}\ \\[-12.5ex]
\begin{quote}
\begin{PPP}{}
\label{newprop}
\ \\[-6.5ex]
\begin{Ventry}{}
\item The family of \co{LS} $\pg{\vert C_{\nh}(\bs{x})}$ satisfies
Properties~\ref{coh}.\\[-4.5ex]
\end{Ventry}
\end{PPP}
\end{quote}
\noindent
\textbf{Proof:}\\[2ex]
\noindent\co{Measurability} and \co{normalization} are straightforward.

\noindent\co{Completeness} can be expressed as
\begin{equation}
\nh \int_{\IT}\mu(\ud\bs{x})\, 
\langle\bs{\ell}\ 
\vert C_{\nh}(\bs{x})\rangle\langle C_{\nh}(\bs{x})\vert
\ \bs{m}\rangle
=
\delta^{(N)}_{\bs{\ell},\bs{m}},\qquad\forall\bs{\ell},\bs{m}\in\ZNZD\ ,
\notag
\end{equation}
where we have introduced the periodic Kronecker
delta, that is 
\mbox{$\delta^{(N)}_{\bs{n},\bs{0}}=1$} if and only if
$\bs{n}\equiv\bs{0}\,\pmod{N}$. This is proved as follows:
\begin{align}
&\phantom{=}
\nh \int_{\IT}\mu(\ud\bs{x})\,  
\langle\,\bs{\ell}\,
\vert C_{\nh}(\bs{x})\,\rangle\langle \,C_{\nh}(\bs{x})\,\vert
\,\bs{m}\,\rangle
= 
\nh 
\int_{0}^{1}\ud x_1\,
\int_{0}^{1}\ud x_2\;
\langle\,\bs{\ell}\, 
\vert 
\,\hat{\bs{x}}_N\,
\rangle\langle 
\,\hat{\bs{x}}_N\,\displaybreak
\vert
\,\bs{m}\,\rangle=\notag\\
&= 
\nh \ 
\delta^{(N)}_{\ell_1\:,\:m_1}\;
\delta^{(N)}_{\ell_2\:,\:m_2}\;\pq{
\int_{0}^{1}\ud x_1\,
\delta^{(N)}_{\ell_1\:,\:\floor{N x_1+\pum}}}\;\pq{
\int_{0}^{1}\ud x_2\;\ 
\delta^{(N)}_{\ell_2\:,\:\floor{N x_2+\pum}}}\;
\notag
\\
&
 = \nh \ 
\pt{\delta^{(N)}_{\ell_1\:,\:m_1}\;
\delta^{(N)}_{\ell_2\:,\:m_2}}\;\pq{
\int_{\frac{\ell_1-\pum}{N}}^{\frac{\ell_1+\pum}{N}}\ud x_1\,
}\;\pq{
\int_{\frac{\ell_2-\pum}{N}}^{\frac{\ell_2+\pum}{N}}\ud x_2}
 = N^2 \ 
\delta^{(N)}_{\bs{\ell},\bs{m}}\;\frac{1}{N^2}
 = \delta^{(N)}_{\bs{\ell},\bs{m}}\ \cdot
\;
\notag
\ 
\end{align}

\co{Localization} comes as follows: 
from Definition~\ref{xxxnnn} (see Remark~\ref{distCS} and
Figure~\ref{lattice}), it turns out that $| 
C_{\nh}(\bs{x})\rangle$ is 
orthogonal to 
every basis element labeled by a point of $L_N$ whose toral
distance $d_{\IT}$ (see Definition~\eqref{dont}) from $\bs{x}$ is
greater than $\frac{1}{N\sqrt{2}}$. 
As a consequence, the quantity $\langle
C_{\nh}(\bs{x}),C_{\nh}(\bs{y})\rangle=0$ 
if the distance on
the torus between $\bs{x}$ and $\bs{y}$ is greater than
$\frac{\sqrt{2}}{N}$.
Thus, given $d_0>0$, it is sufficient to choose
$N_0(\epsilon,d_0)>\sqrt{2}/d_0$, to have 
\begin{equation*}
N>N_0(\epsilon,d_0) \Longrightarrow \nh\;\langle
C_{\nh}(\bs{x}),C_{\nh}(\bs{y})\rangle=0\ \cdot\hfill\tag*{\qed}
\end{equation*}\ \\[-7.5ex]
\begin{quote}
\begin{NNS}{}\ \\[-5.5ex]
\begin{Ventry}{2)}
\item[(1)] The last result in the previous Proposition amounts to
an even stronger
\co{localization} property than property~\ref{coh}.4; this is due
to our particular choice of \co{L}attice \co{S}tates, which, as we
shall see, is suited to the task of controlling \co{S}awtooth
\co{M}aps. In general, one can hardly hope to achieve orthogonality
and must be content with the weaker \co{localization} condition~\ref{coh}.4.
\item[(2)] Although the set of \co{LS} of Definition~\ref{xxxnnn}
fulfill Properties~\ref{coh}, which are typical of \co{C}oherent
\co{S}tates, \co{LS} differ from them in that the context we are
considering is commutative. In spite of this, it is convenient to
adopt the formalism of Quantum
Mechanics; in particular the set of
\co{LS} is interpreted as a Hilbert orthonormal basis of Dirac kets,
whose corresponding projectors
form a partition of unit into indicator functions
having support on small squares of the torus, as in
Figure~\ref{lattice}, whose sides scales as $\frac{1}{N}\;\cdot$
\end{Ventry}
\end{NNS}
\end{quote}
\subsection{\co{A}nti--\co{W}ick Discretization and its  continuous
limit on $\IT$} 
\label{AWD}
\vspace{3mm}  
In order to study the continuous limit and, more generally,
the quasi--continuous behaviour of $\tripQS$
when $N\to\infty$, we follow the
semi--classical technique known as \co{A}nti--\co{W}ick
quantization. The other standard quantization technique, namely
the \co{W}eyl procedure, despite being more straightforward and less
technically heavy, is nevertheless more suited to smooth spaces of
functions and was indeed instrumental in the study of discretized Cat
Maps~\cite{Ben04:1}.
Instead, in our case, the
\co{A}nti--\co{W}ick procedure is a better choice due to the
discontinuous character of the dynamics, as it will clearly appear in
the next Section.   

\noindent We start choosing concrete
discretization/de--discretization *-morphisms.\\[-2ex]
\begin{quote}
\begin{DDS}{}
\label{qWick}
\ \\
 Given the family $\{\vert C_{\nh}(\bs{x})\rangle\}$ of
 \co{L}attice \co{S}tates in 
 ${\cal H}_{\nh}$, the \co{A}nti-\co{W}ick--like
 discretization scheme (\co{AW}, for short) will be described by a one
 parameter family of
 (completely) positive unital map ${\cal J}_{\nh,\infty}:
 \Lspace{\infty}{\IT}\to\c 
 D_{\nh}$ 
 \begin{equation*}
 {\Lspace{\infty}{\IT}\ni} f \mapsto   
 \nh \int_{{\IT}}\mu(\ud \bs{x})\, f(\bs{x})\,
  \coh{\nh}{\bs{x}} \lcoh{\nh}{\bs{x}}=:{\cal J}_{\nh,\infty}(f)\in \cal D_{\nh}\quad .
 \end{equation*}
 The corresponding de--discretization operation will be described by 
 the (completely) positive unital map ${\cal J}_{\infty,\nh}: {\cal
 D}_{\nh} 
 \to\Lspace{\infty}{\IT}$ 
 \begin{equation*}
  {\cal D}_{\nh} \ni X \mapsto 
  \<C_{\nh}(\bs{x}), X\,C_{\nh}(\bs{x})\>=:{\cal J}_{\infty,\nh}(X)(\bs{x})
 \in\Lspace{\infty}{\IT}
  \quad .
 \end{equation*}\\[-7.5ex]
\end{DDS}
\begin{NNS}{}\ \\[-5.5ex]
\label{tauenneonD}
\begin{Ventry}{\mdseries ii.}
\item[\mdseries i.] Both maps are identity preserving (unital) because
of the conditions satisfied 
by the family of \co{L}attice \co{S}tates and are completely positive,
since both $\Lspace{\infty}{\IT}$ and $\cal D_{\nh}$ are commutative algebras.
One can also check that:
\begin{equation}
 \norm{{\cal J}_{\infty,\nh} \circ {\cal J}_{\nh,\infty}(g)}{\infty} \le
 \norm{g}{\infty},\quad g\in \Lspace{\infty}{\IT}\ \cdot
\notag
\end{equation}
\item[\mdseries ii.] Definition~\ref{qWick} yields
$\tn\circ{\cal 
J}_{\nh,\infty}=\omega_\mu$, with $\tn$ given in~\eqref{tauenne}.
\end{Ventry}
\end{NNS}
\end{quote}
In Appendix~\ref{AWPSCS}, more operative details are presented, whereas
in the following we prove some simple properties that 
incorporate minimal requests for rigorously defining the sense in
which the discrete dynamical systems 
$\tripQS$ tends to $\tripAS$, when $\frac{1}{N} \to 0$. 
\begin{quote}
\begin{PPP}{}
\label{prop1}
\ \\[-5.5ex]
\begin{Ventry}{3}
\item[(1)] 
 For all $f\in\Lspace{\infty}{\IT}$ and $X\in {\cal D}_\nh$,
 \begin{equation*}
\omega_\mu\pt{\overline{g}\  {\cal J}_{\infty,\nh}\pt{X}} = \tn \bigl( {\cal
J}_{\nh,\infty}(g)^* X\bigr)\ ;
 \end{equation*}
\item[(2)] 
 For all $f,g\in\Lspace{\infty}{\IT}$
 \begin{equation*}
  \lim_{N\to\infty} \tn \bigl( {\cal J}_{\nh,\infty}(f)^*
  {\cal J}_{\nh,\infty}(g) \bigr) = \omega_\mu(\overline f g) = \int_{\IT}
  \mu(\ud\bs{x})\, \overline{f(\bs{x})}g(\bs{x}).
 \end{equation*}
\item[(3)] For all $X\in {\cal D}_\nh$, and for all $N\in\IN^+$,
 \begin{equation*}
  {\cal J}_{\nh,\infty}\circ {\cal J}_{\infty,\nh} \ (X)= X\ ;
 \end{equation*}
\item[(4)] 
 For all $f\in\Lspace{\infty}{\IT}$
 \begin{equation*}
  \lim_{N\to\infty} {\cal J}_{\infty,\nh} \circ {\cal J}_{\nh,\infty}(f) =
  f\quad \mu\text{ -- a.e.}
 \end{equation*}
\end{Ventry}
\end{PPP}
\end{quote}
\noindent
\textbf{Proof:}\\[2ex]
The first two statements in the above Proposition directly follow
from Definitions~\ref{qWick} together with~\eqref{CSforL1}; the latter
two are equivalent and their proof can be found
in~\cite{Ben03:1}, the only difference being the dimension
$\nh$ of the Hilbert space $\cal H_\nh$, here $\nh=N^2$, there
$\nh=N^{\phantom{2}}\!\!$.\hfill$\qed$
\begin{quote}\ \\[-7.5ex]
\begin{NNN}{}\ \\[-5.5ex]
\begin{Ventry}{}
\item Properties 1 and 2 in the previous Proposition show how (GNS)
scalar products in the discrete, respectively continuous limit, are
related; properties 3 and 4 concern instead the direct--inverse
relations between the discretization and the de--discretization maps.
\end{Ventry}
\end{NNN}
\end{quote}
\section{Discretization of the Dynamics}
\label{QOD}
\vspace{6mm}
\subsection{Classical description of \co{S}awtooth
\co{M}aps}
\label{AATT}
\vspace{3mm}  
We shall now focus on a special class of automorphisms of the torus,
namely the \co{S}awtooth \co{M}aps~\cite{Che92:1,Vai92:1} (\co{SM} for
short), that is on triples $(\IT,\mu,S_\alpha)$ where
\begin{align}
S_\alpha 
\begin{pmatrix}
x_1\\
x_2
\end{pmatrix} & =
\begin{pmatrix}
1+\alpha & 1\\
\alpha & 1
\end{pmatrix}
\begin{pmatrix}
\bk{x_1}\\
x_2
\end{pmatrix}
\ \pmod{1}\ ,\quad
\alpha\in\IR
\label{AoDC_1}
\\
& = \begin{pmatrix}
\bk{\pt{1+\alpha} \bk{x_1} +  x_2}\\
\bk{\alpha \bk{x_1} + x_2}
\end{pmatrix}\nonumber
\end{align} 
\begin{quote}
\begin{NNS}{}\ \\[-6.5ex]
\label{Rem_31}
\begin{Ventry}{\mdseries vii.}
\item[\mdseries i.] In the following, a point $\bs{x}$ of the torus,
will correspond to an 
equivalence class of $\IR^2$ points whose coordinates differ by
integer values;
\item[\mdseries ii.] without the fractional part,~\eqref{AoDC_1} is
not well defined on $\IT$ for 
not--integer $\alpha$; indeed,
the same point $\bs{x} = \bs{x} + \bs{n} \in \IT , \bs{n} \in
{\IZ}^2$, would have (in general) $S_\alpha \pt{\bs{x}} \neq S_\alpha
\pt{\bs{x}+\bs{n}}$. Of course, $\bk{\cdot}$ is not necessary when
$\alpha\in{\IZ}$;
\item[\mdseries iii.] the Lebesgue measure on $\IT$ is
\enfasi{invariant} for all $\alpha\in\IR$;
\item[\mdseries iv.] if $\alpha\not\in\IZ$, the $S_\alpha$ are known
as \co{S}awtooth \co{M}aps;
\item[\mdseries v.] when $\alpha\in\IZ$,
	we shall write $T_\alpha$ instead of
	$S_\alpha$.
	$T_1=\pt{\begin{smallmatrix} 2 & 1\\ 1 & 1
	\end{smallmatrix}}$ is the Arnold Cat Map~\cite{Kat99:1}.
	In general, $\displaystyle T_1 \in 
	{\left\{T_{\alpha}\right\}}_{\alpha\in\IZ} \subset
	{\text{SL}}_2\left(\IZ\right)\subset
	{\text{GL}}_2\left(\IZ\right)\subset
	{\text{M}}_2\left(\IZ\right)$ where
	${\text{M}}_2\left(\IZ\right)$ is the subset of $2\times2$
	matrices with integer entries, 
	${\text{GL}}_2\left(\IZ\right)$ the subset of
	invertible matrices and ${\text{SL}}_2\left(\IZ\right)$ the subset
	of matrices with determinant one:
	the dynamics generated by
	$T_{\alpha}\in{\text{SL}}_2\left(\IZ\right)$
	is called \enfasi{Unimodular Group}~\cite{Kat99:1} (\co{UMG}
	for short);
\item[\mdseries vi.] after identifying $\bs{x}$ with canonical coordinates
$(q,p)$ and imposing the$\pmod{1}$ condition on
both of them, the above dynamics reads
\begin{equation}
\begin{cases}
q^\prime &= q + p^\prime\\
p^\prime &= p + \alpha  \bk{q}
\end{cases}
\pmod{1}\ \cdot
\label{AoDC_11}
\end{equation}
This is nothing but the Chirikov Standard Map~\cite{Cas95:1} in which
$-\frac{1}{2\pi}\sin(2\pi q)$ is replaced by 
$\bk{q}$.
The dynamics in~\eqref{AoDC_11} can also be thought of as generated
by the (singular) Hamiltonian 
\begin{equation}
H(q,p,t)=\frac{p^2}{2}-\alpha\, 
\frac{{\bk{q}}^2}{2}\,\delta_p(t),
\nonumber
\end{equation}
where $\delta_p(t)$ is the periodic Dirac delta which makes the
potential act through
periodic kicks with period 
$1$~\cite{For91:1};
\item[\mdseries vii.] \co{S}awtooth \co{M}aps are invertible and the
inverse is given by the expression   
\begin{align}
S_\alpha^{-1}
\begin{pmatrix}
x_1\\
x_2
\end{pmatrix} & =
\begin{pmatrix}
\phantom{-}1 & 0\\
-\alpha & 1
\end{pmatrix}
\bk{\begin{pmatrix}
1 & -1\\
0 & \phantom{-}1
\end{pmatrix}
\begin{pmatrix}
x_1\\
x_2
\end{pmatrix}}
\ \pmod{1}
\label{AoDC_1d}
\\
& =
\begin{pmatrix}
\bk{x_1 - x_2}\\
\bk{\bk{x_2} - \alpha\bk{x_1 - x_2}}
\end{pmatrix}
\nonumber
\intertext{or, in other words,}
& 
\begin{cases}
q &= \phantom{-\alpha\,}q^\prime - p^\prime\\
p &= -\alpha\,q^{\phantom{\prime}} + p^\prime
\end{cases}
\pmod{1}\ .
\notag
\end{align}
It can indeed be checked that
$S_\alpha\pt{S_\alpha^{-1}\pt{\bs{x}}} =
S_\alpha^{-1}\pt{S_\alpha\pt{\bs{x}}}=\bs{x},\ \forall
\bs{x}\in\IT$.\\
Further, $S_\alpha^{-1}$ preserves the Lebesgue measure on $\IT$.
\end{Ventry}
\end{NNS}
\end{quote}
We now list a set of properties~\cite{Che92:1,Vai92:1,Per87:1} of
\co{S}awtooth \co{M}aps that will be used in the following
\begin{quote}
\begin{PRS}[of \co{S}awtooth \co{M}aps]\ \\[-6.5ex]
\label{Pro_31}
\begin{Ventry}{4}
\item[(1)] \co{S}awtooth \co{M}aps $\{S_\alpha\}$ are
	\enfasi{discontinuous} on
	the subset \\$\gamma_0\coleq\pg{\bs{x} = \pt{0,p},\
	p\in{\mathds{T}}}\in \IT$: 
	two points close to $\gamma_0$,
	\mbox{$A\coleq\pt{\varepsilon,p}$} and
	$B\coleq\pt{1-\varepsilon,p}$,
	have images that differ, in the $\varepsilon \rightarrow 0$ limit, by a
	vector
$d^{(1)}_{S_{\alpha}^{\phantom{-1}}}(A,B)=\pt{\alpha,\alpha}
\pmod{1}$. 
\item[(2)] Inverse \co{S}awtooth \co{M}aps $\{S_\alpha^{-1}\}$ are
	\enfasi{discontinuous} on
	the subset \\$\gamma_{-1}\coleq S_\alpha\pt{\gamma_0} =
	\pg{\bs{x} = \pt{p,p},\
	p\in{\mathds{T}}}\in \IT$:
	two points close to $\gamma_{-1}$, namely
	\mbox{$A\coleq\pt{p+\varepsilon,p-\varepsilon}$} and
	$B\coleq\pt{p-\varepsilon,p+\varepsilon}$,
	have images that differ, in the $\varepsilon \rightarrow 0$ limit, by a
	vector $d^{(1)}_{S_{\alpha}^{-1}}(A,B)=\pt{0,\alpha} \pmod{1}$.
\item[(3)] The maps $T_\alpha$ and $T_\alpha^{-1}$ are
	\enfasi{continuous}: \\
	$\alpha\in\IZ \Longrightarrow
	d^{(1)}_{T_{\alpha}^{\phantom{-1}}}(A,B) =
	d^{(1)}_{T_{\alpha}^{-1}}(A,B) = \pt{0,0} \pmod{1}$.  
\item[(4)] The eigenvalues of the matrix 
	$S_\alpha=\pt{\begin{smallmatrix} 1+ \alpha  & 1\\ \alpha & 1 
	\end{smallmatrix}}$ are  
 	$\pt{\alpha+2\pm\sqrt{(\alpha+2)^2-4}}/2$.
	They are conjugate complex
	numbers if
	$\alpha\in\pq{-4,0}$, whereas one eigenvalue $\lambda>1$ if 
	$\alpha\not\in\pq{-4,0}$.
	In this case,
	distances are stretched along the direction of the
	eigenvector $|\bs{e}_+\rangle$,
	$S_\alpha|\bs{e}_+\rangle=\lambda|\bs{e}_+\rangle$,
	contracted along 
	that of $|\bs{e}_-\rangle$, $S_\alpha|\bs{e}_-\rangle=
	\lambda^{-1}|\bs{e}_-\rangle$:
	$\log\lambda$ is a (positive) \enfasi{Lyapunov exponent}.\\
	For such $\alpha$'s all periodic points are
	hyperbolic~\cite{Per87:1}.	
\end{Ventry}
\end{PRS}
\begin{figure}[H]
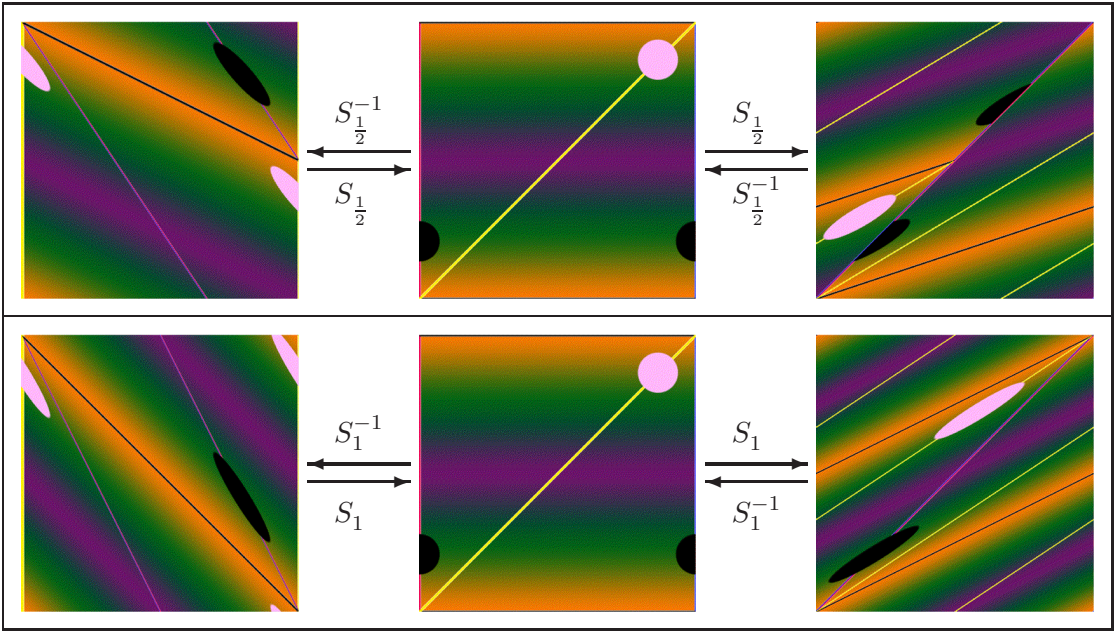

\begin{center}
\begin{picture}(248,140)(-4,-4) 

\put(-4,-4){\framebox(248,70){}}
\put(-4,-4){\framebox(248,140){}}

\put(0,70){\includegraphics[width=62\unitlength,height=62\unitlength]{./inv_000_436.epsi}}
\put(89,70){\includegraphics[width=62\unitlength,height=62\unitlength]{./ori_001_436.epsi}}
\put(178,70){\includegraphics[width=62\unitlength,height=62\unitlength]{./dir_002_436.epsi}}

\thicklines

\put(87,103){\vector(-1,0){23}}   
\put(64,99){\vector(1,0){23}}    
\put(153,103){\vector(1,0){23}}   
\put(176,99){\vector(-1,0){23}}   
\thinlines

\put(64,103){\makebox(23,14)[cc]{$S_{\frac{1}{2}}^{-1}$}}
\put(153,103){\makebox(23,14)[cc]{$S_{\frac{1}{2}}^{\phantom{-1}}$}}
\put(64,85){\makebox(23,14)[cc]{$S_{\frac{1}{2}}^{\phantom{-1}}$}}
\put(153,85){\makebox(23,14)[cc]{$S_{\frac{1}{2}}^{-1}$}}

\put(0,0){\includegraphics[width=62\unitlength,height=62\unitlength]{./inv_003_436.epsi}}
\put(89,0){\includegraphics[width=62\unitlength,height=62\unitlength]{./ori_001_436.epsi}}
\put(178,0){\includegraphics[width=62\unitlength,height=62\unitlength]{./dir_005_436.epsi}}

\thicklines

\put(87,33){\vector(-1,0){23}}   
\put(64,29){\vector(1,0){23}}    
\put(153,33){\vector(1,0){23}}   
\put(176,29){\vector(-1,0){23}}   
\thinlines

\put(64,33){\makebox(23,14)[cc]{$S_{1}^{-1}$}}
\put(153,33){\makebox(23,14)[cc]{$S_{1}^{\phantom{-1}}$}}
\put(64,15){\makebox(23,14)[cc]{$S_{1}^{\phantom{-1}}$}}
\put(153,15){\makebox(23,14)[cc]{$S_{1}^{-1}$}}

\end{picture}
\caption[]{
In the upper row, we depict the effects of the
discontinuities of a \co{SM} with $\alpha=\frac{1}{2}$; the picture in
the middle shows the discontinuity lines $\gamma_0$ and $\gamma_{-1}$,
whereas those on the right and left show how they evolve
backward and forward in time. The different parallel bands help
the reader to figure out the toral periodicity and the discontinuous
character of the map, also highlighted by the aperiodic splits of two
spots. Further, for sake of comparison, the lower row presents the
same case of the upper one but for the continuous dynamics ($\alpha=1$).}
\ 
\label{classSM}
\end{center}
\end{figure}\ \\[-9ex]
\begin{NNS}{}\ \\[-2.5ex]
Because of the presence of the fractional part
in~\eqref{AoDC_1}~and~\eqref{AoDC_1d}, we have to 
distinguish the action of $S_\alpha$ and $S_\alpha^{-1}$ from a mere
matrix action. We shall adopt the following notations:
\begin{Ventry}{\mdseries iii.}
\item[\mdseries i.] With $S_\alpha$ the matrix $\pt{\begin{smallmatrix} 1+ \alpha  & 1\\
\alpha & 1 \end{smallmatrix}}$ in Property~\ref{Pro_31}.4, the expression
$S_\alpha\pt{\bs{x}}$ will denote the action represented
by~\eqref{AoDC_1}, whereas $S_\alpha\cdot \bs{x}$ will denote the
matrix action of $S_\alpha$ on the vector $\bs{x}$.
\item[\mdseries ii.] When the dynamics arises from the action of the
\co{UMG} (see Remark~\ref{Rem_31}.v.), so, in
particular, when ${\left\{T_{\alpha}\right\}}_{\alpha\in\IZ}$
is the family of toral automorphisms, equation~\eqref{AoDC_1} assumes
the simpler form 
$T_\alpha \pt{\bs{x}}= T_\alpha\cdot \bs{x}
\ \pmod{1}$.
\item[\mdseries iii.] Analogously, expressions like $T_\alpha \cdot\bs{x},\
T_\alpha^{\text{tr}} \cdot\bs{x},\ T_\alpha^{-1} \cdot\bs{x}$
and $\pt{T_\alpha^{\text{tr}}}^{-1}\cdot\bs{x}$, will denote the actions by
$T_\alpha$ itself, its transposed, its
inverse and the inverse of the transposed, respectively.
\end{Ventry}
\end{NNS}
\end{quote}\ \\[-10.5ex]
\subsection{Algebraic description of continuous and discretized \co{S}awtooth
\co{M}aps}
\label{AdoT}
In this Section we make use of the commutative (Von~Neumann) algebra
$\Lspace{\infty}{\IT}$ introduced in
Section~\ref{CDS} and consider the algebraic description of
\co{S}awtooth \co{M}aps by triples $\tripASa$, where $\om$ has been
defined in~\eqref{omegamu} and $\Theta_{\alpha}: \Lspace{\infty}{\IT}\mapsto 
\Lspace{\infty}{\IT}$ is the discrete--time dynamics 
generated as follows:  
\begin{equation}
\Theta_{\alpha}\pt{f}\pt{\bs{x}}\coleq f(S_{\alpha}
\pt{\bs{x}})\quad,\quad \alpha\in\IR\ \cdot 
\notag
\end{equation}
The maps $\Theta_{\alpha}^j,\ j\in\IZ$ are automorphisms of
$\Lspace{\infty}{\IT}$ and leave the 
state $\omega_\mu$ invariant.

Our aim is now to define a suitable discrete evolution
$\Theta_{\nh,\alpha}$ on ${\cal D}_\nh$, such that the discretized
triplets $\tripQSa$ converge to the continuous \co{SM}.

We start by introducing two different kinds of maps: the first ones,
$U_{\alpha}^{\pm j}, \ j\in\IZ$, are defined on
the torus $\IT\pt{{[0,N)}^2}$, namely $[0,N)\times [0,N) \pmod{N}$, and
given by
\begin{subequations}
\label{Ualpha}
\begin{alignat}{2}
\IT\pt{{[0,N)}^2} \ni\bs{x}&\mapsto &
U_\alpha^0\pt{\bs{x}} &\coleq 
\bs{x}
\notag\\
&&&= N\,S_\alpha^0\pt{\frac{\bs{x}}{N}}
\in\IT\pt{{[0,N)}^2}\ ,
\label{Ualphaa}
\\
\IT\pt{{[0,N)}^2} \ni\bs{x}&\mapsto &
U_\alpha^{\pm 1}\pt{\bs{x}} &\coleq N\,
S_\alpha^{\pm 1}\pt{\frac{\bs{x}}{N}}
\in\IT\pt{{[0,N)}^2}\ ,
\label{Ualphab}
\\
\IT\pt{{[0,N)}^2} \ni\bs{x}&\mapsto& 
U_{\alpha}^{\pm j}\pt{\bs{x}}&\coleq
\underbrace{
U_{\alpha}^{\pm 1}(\,
U_{\alpha}^{\pm 1}(\,
\cdots
U_{\alpha}^{\pm 1}(\,
U_{\alpha}^{\pm 1}(
}_{j\ \text{times}}
\bs{x}\,)\,)\cdots \,)\,)\ ,
\quad j\in{\IN}^{+}\ ,\notag\\
&&&= N\,
S_\alpha^{\pm j}\pt{\frac{\bs{x}}{N}}
\in\IT\pt{{[0,N)}^2}\ \cdot
\label{Ualphac}
\end{alignat}
\end{subequations}
The second class consists of maps $V_\alpha^{\pm j}$ from
$\IT\pt{{[0,N)}^2}$ onto
its subset ${\pt{\IZ / N \IZ}}^2$, whose actions are as follows
\begin{subequations}
\label{Valpha}
\begin{alignat}{2}
\IT\pt{{[0,N)}^2} \ni\bs{x}&\mapsto& 
V_\alpha^0\pt{\bs{x}} &\coleq 
\floor{\bs{x}}
\notag\\
&&&= \pm\floor{\pm \,U_\alpha^{0}\pt{\floor{\bs{x}}}}\in
\ZNZD
\ ,
\label{Valphaa}
\\
\IT\pt{{[0,N)}^2} \ni\bs{x}&\mapsto &
V_\alpha^{\pm 1}\pt{\bs{x}} &\coleq 
\pm\floor{\pm \,U_\alpha^{\pm 1}\pt{\floor{\bs{x}}}}\in
\ZNZD
\ ,
\label{Valphab}
\\
\IT\pt{{[0,N)}^2} \ni\bs{x}&\mapsto &
\:V_{\alpha}^{\pm j}\pt{\bs{x}}&\coleq
\underbrace{
V_{\alpha}^{\pm 1}(\,
V_{\alpha}^{\pm 1}(\,
\cdots
V_{\alpha}^{\pm 1}(\,
V_{\alpha}^{\pm 1}(
}_{j\ \text{times}}
\floor{\bs{x}}\,)\,)\cdots \,)\,)\ ,
\quad j\in{\IN}^{+}\ ,\notag
\end{alignat}
\begin{equation}
\ \ \ \ \ \ =\underbrace{
\pm\lfloor \pm \,U_{\alpha}^{\pm 1}(\,
\pm\lfloor \pm \,U_{\alpha}^{\pm 1}(\,
\cdots
\pm\lfloor \pm \,U_{\alpha}^{\pm 1}(\,
\pm\lfloor \pm \,U_{\alpha}^{\pm 1}(
}_{j\ \text{times}}
\floor{\bs{x}}\,)\rfloor\,)\rfloor\cdots \,)\rfloor\,)\rfloor\in
\ZNZD\ \cdot
\label{Valphac}
\end{equation}
\end{subequations}
\begin{quote}\ \\[-7.5ex]
\begin{NNN}{}\ \\[-5.5ex]
\begin{Ventry}{}\label{percheUV}
\item The maps $U_\alpha^j$ are extensions of the $S_\alpha^j$
on the enlarged torus $\IT\pt{{[0,N)}^2}$; however, they do not map
the lattice $L_N$ into itself, therefore we are forced to use the maps
$V_\alpha^j$ to define a consistent discretized dynamics. 
\end{Ventry}\ \\[-5.5ex]
\end{NNN}
\begin{DDD}{}\ \\[-5.5ex]
\label{ThetaNSa}
	\item[$\Theta_{\nh,\alpha}$] will denote the map:
	\begin{equation}
	{\cal D}_\nh\ni X \mapsto
	\Theta_{\nh,\alpha}^{\phantom{t}}\pt{X}  
	\coleq\sum_{\bs{\ell} \in {(\ZNZ{N})^2}} 
	X_{V_\alpha\pt{\bs{\ell}},V_\alpha\pt{\bs{\ell}}}
	\ket{\bs{\ell}}\bra{\bs{\ell}}\in{\cal D}_\nh
	\ \cdot 
\label{CoAFE_11}
	\end{equation}
\end{DDD}
\end{quote}
$\Theta_{\nh,\alpha}^{\phantom{t}}$ is a
*-automorphism of $\c D_\nh$; indeed, the map
\begin{equation*}
{\pt{\IZ / N \IZ}}^2\ni\bs{\ell}\longmapsto
V_\alpha\pt{\bs{\ell}}\in{\pt{\IZ / N \IZ}}^2
\end{equation*}
is a
bijection, so that~\eqref{CoAFE_11} can be rewritten in the
more convenient form 
\begin{align}
\Theta_{\nh,\alpha}^{\phantom{t}}\pt{X} & 
= \sum_{\bs{\ell} \in {(\ZNZ{N})^2}} 
X_{V_\alpha\pt{\bs{\ell}},V_\alpha\pt{\bs{\ell}}}
\ket{\bs{\ell}}\bra{\bs{\ell}}=
\notag
\\
& = \sum_{V_\alpha^{-1}\pt{\bs{s}} \in {(\ZNZ{N})^2}} 
X_{\bs{s},\bs{s}}
\ket{V_\alpha^{-1}\pt{\bs{s}}}\bra{V_\alpha^{-1}\pt{\bs{s}}}=
\notag\\
\!\!\!\!\!\!\!\!\!{\textstyle\text{(see Remark~\ref{def_Vj}.{\mdseries
iii.} below)}}\ \ 
& = W_{\alpha,N}^{\phantom{*}}\pt{\sum_{\newatop{\text{all
equiv.}}{\text{classes}}}  
X_{\bs{s},\bs{s}}
\ket{\bs{s}}\bra{\bs{s}}}W_{\alpha,N}^{*}= 
\label{CoAFE_121}
\\
& = W_{\alpha,N}^{\phantom{*}}
\;X\;\;
W_{\alpha,N}^{*}\ ,
\notag
\end{align}
where the operators $W_{\alpha,N}$, defined by linearly extending
the maps
\begin{equation}
{\cal H}_\nh\ni\big|\bs{\ell}\big\rangle\longmapsto
W_{\alpha,N}^{\phantom{*}}\big|\bs{\ell}\big\rangle\coleq\ket{
V_\alpha^{-1}\pt{\bs{\ell}}}\in{\cal H}_\nh\ \cdot
\label{aggiunta}
\end{equation} 
to $\c H_\nh$,
are unitary: $\displaystyle W_{\alpha,N}^*\big|\bs{\ell}\big\rangle\coleq\ket{
V_\alpha\pt{\bs{\ell}}}$.

\noindent For the same reason the state $\tn$ is
$\Theta_{\nh,\alpha}^{\phantom{t}}$--invariant and $V_\alpha$ is
invertible too.

\noindent Note that $\Theta_{\nh,\alpha}^{j}\coleq
\underbrace{\Theta_{\nh,\alpha}^{\phantom{t}}\circ  
\cdots\circ
\Theta_{\nh,\alpha}^{\phantom{t}}}_{j\ \text{times}}$ is implemented by
$V_{\alpha}^{j}\pt{\bs{\ell}}$ given
in~\eqref{Valphac}. 

\begin{quote}
\begin{NNS}{}\ \\[-5.5ex]
\label{def_Vj}
\begin{Ventry}{\mdseries iii.}
\item[\mdseries i.] The double $\pm$ sign in front and within every
floor function in
equations~\eqref{Valpha} is
needed in order to have $V_{\alpha}^{\pm j}(\,
V_{\alpha}^{\mp j}(\bs{x}))=V_{\alpha}^{0}(\bs{x})$ (the identity
when $\bs{x}\in{\pt{\IZ / N \IZ}}^2$); the reason is
that, in general, $\floor{-x}\neq-\floor{x}$, for $x\not\in\IZ$ (see~\cite{Apo76:1}).
\item[\mdseries ii.] When $\alpha\in\IZ$,
${\pt{\IZ / N \IZ}}^2\ni\bs{\ell}\longmapsto V_\alpha\pt{\bs{\ell}} =
T_\alpha \cdot\bs{\ell}\in{\pt{\IZ / N \IZ}}^2$, namely the action of the
map $V_{\alpha}$ becomes 
that of a matrix$\pmod{N}$. Moreover, in that case, $U_\alpha$ and
$V_\alpha$ coincide.
\item[\mdseries iii.] Since 
	$\bs{\ell}\longmapsto V_\alpha\pt{\bs{\ell}}$ 
	is a bijection, 
	in~\eqref{CoAFE_121} one can sum over the equivalence classes.
\end{Ventry}
\end{NNS}
\end{quote}
\noindent
\section{Continuous limit of the dynamics}
\label{CCLD}
\vspace{6mm}
One of the main issues in the semi-classical analysis is to compare
if and how the quantum and classical time evolutions mimic each other
when a suitable quantization parameter goes to zero.

In this article we are instead considering the possible agreement between
the dynamics of continuous classical systems and that of a class of
discrete approximants.
In practice, in our case, we will study the difference 
\begin{equation}
\Theta_\alpha^j - {\cal J}_{\infty,\nh}\circ\Theta_ {\nh,\alpha}^j\circ
{\cal J}_{\nh,\infty}
\label{convergenza}
\end{equation}
which represents how much the discrete dynamics at timestep $j$
differs from the continuous one at the same timestep.

For quantum systems, whose classical limit is chaotic, the situation is
strikingly different from those with regular classical limit.
In the former case, classical and quantum mechanics agree, that is a
difference as in~\eqref{convergenza} is negligible,
only over times $j$ which scale logarithmically (and not as a power
law) in the quantization
parameter.  

\noindent As we shall see, such a type of scaling is not exclusively
related with 
non--commutativity; in fact, the quantization--like procedure
developed so 
far, exhibits a similar behavior when $N\to\infty$ and we
recover $\tripASa$ as a continuous limit of $\tripQSa$.
\subsection{Continuous limit for \co{S}awtooth \co{M}aps}
\label{ClfSM}
\vspace{3mm} 
Later on we shall show that the difference in~\eqref{convergenza} goes to
zero in a suitable topology; for the moment we just note that the
major difficulties in the proof are due to the 
discontinuous character of the fractional part that appears
in~\eqref{AoDC_1}.

\noindent It is therefore important to briefly discuss the discontinuities of
the maps $S_\alpha$~\cite{Che92:1,Vai92:1,Per87:1}.

\noindent
As already noted in Property~\ref{Pro_31}.1, $S_\alpha$ is
discontinuous 
on the circle $\gamma_0$;
therefore $S_\alpha^n$ will be discontinuous on the preimages 
\begin{subequations} 
\label{gammakk}
\begin{alignat}{3}
\gamma_m & \coleq S_\alpha^{-m}\pt{\gamma_0} & \ & \text{ for }& \ 0 & \leq
m<n
\label{gammakka}
\ ,  
\intertext{whereas the discontinuities of $S_\alpha^{-n}$ lie on the sets} 
\gamma_{-m} & \coleq S_\alpha^{\,m}\pt{\gamma_0} & \ & \text{ for }& \ 0
& < m \leq n\
\cdot 
\label{gammakkb}
\end{alignat}
\end{subequations} 
\noindent
Apart from $\gamma_{-1}$, whose projection on the $[0,1)^2$ square is
its diagonal (see Fig.~\ref{tre}), each set of the type
$\gamma_m$ (for $\gamma_{-m}$ the argument is similar) is the
(disjoint) union of segments parallel to each other 
whose endpoints lie either on the same segment belonging to
$\gamma_p$, $p<m$, or on two different segments belonging to
$\gamma_p$ and $\gamma_{p^\prime}$, with $p^\prime\leq p <
m$~\cite{Vai92:1}.\\
It proves convenient to introduce the
\enfasi{discontinuity set} of
$S_\alpha^n$,
\begin{equation}
\IT \supset \Gamma_n\coleq \bigcup_{p=0}^{n-1} \; \gamma_p
\ ,
\label{Gamman}
\end{equation}
and its
complementary set, $G_n\coleq \IT \setminus \Gamma_n$.

\noindent We now enlarge the previous definition from 
continuous \co{S}awtooth \co{M}aps, to discretized ones.\\[-2ex]
\begin{quote}
\begin{DDS}{}\ \\
\label{Gnbar}
We shall call ``segment'', and denote it by $\pt{A,B}$, the shortest
curve joining $A,B\in\IT$, by $l\pt{{\gamma}_p}$ the length of the
curve ${\gamma}_p$ and by
\begin{align}
\overline{\gamma}_p\pt{\varepsilon} &\coleq 
\Big\{
\bs{x}\in\IT
\ \Big|\ 
d_{\IT}\pt{\bs{x},\gamma_p} \leq \varepsilon
\Big\}
\label{Gnbar_0}
\intertext{the strip around $\gamma_p$ of width $\varepsilon$, where the distance $d_{\IT}\pt{\cdot,\cdot}$ on the torus
has been introduced in Definition~\ref{dont}.}
\intertext{Further, we shall denote by}
\overline{\Gamma}_n\pt{\varepsilon} &\coleq 
\bigcup_{p=0}^{n-1}\;
\overline{\gamma}_p\pt{\varepsilon}
\label{Gnbar_1}
\intertext{the union of the strips up to $p=n-1$ and by
${G_n^N}\pt{\varepsilon}$
the subset of points}
{G_n^N}\pt{\varepsilon}&\coleq
\pg{\bs{x}\in\IT\ \Big\vert\
\frac{\hat{\bs{x}}_N}{N}\not\in\overline{\Gamma}_n\pt{\varepsilon}} \ ,
\label{Gnbar_2}
\end{align}
where the lattice points ${\hat{\bs{x}}_N}$ have been
introduced in Definition~\ref{xxxnnn}.
\end{DDS}
\end{quote}
\noindent
As already observed, in order to prove that the discretized \co{SM}
tend to continuous \co{SM} 
when $N\to\infty$, the main problem is to
control the discontinuities. It proves convenient to subdivide
the lattice points in a \enfasi{good} and a \enfasi{bad} set and 
show that, on the former, $V_\alpha^q\simeq U_\alpha^q$, at least
on a certain time--scale (see Remark~\ref{percheUV}). This will not turn out to be true for the bad
set, however we shall show that the latter tends with $N$ to a set of zero
Lebesgue measure and thus becomes ineffective.

Following this strategy, we shall concretely show that the
difference~\eqref{convergenza} goes to zero 
with $N \to\infty$ in the strong topology over the Hilbert space
 $\Lspace{2}{\IT}$. More precisely, we have the following theorem
\begin{quote}
\begin{TT}{}\ \\[2ex]
\label{propval}
 Let $\tripQSa$ be a sequence of discretized \co{SM} as defined in
 Section~\ref{QOD}: for all $\gamma>3$,  
 \begin{equation}
\forall f\in \Al\quad,\quad  \slim_{\substack{j,N\to\infty\\
j<\frac{1}{\gamma}\frac{\log N}{\log
\eta}}} 
 \pt{\Theta_\alpha^j- {\cal J}_{\infty,\nh}\circ\Theta_ {\nh,\alpha}^j\circ
{\cal J}_{\nh,\infty}}\pt{f}
= 0\quad ,
\label{added2}
\end{equation}
where the limit is in the strong topology over the Hilbert space
 $\Lspace{2}{\IT}$ and $\eta>\sqrt{2}$ is the largest eigenvalue of
 the matrix $\abs{S_\alpha}\coleq\sqrt{S_\alpha^{\dagger}S_\alpha^{\phantom{\dagger}}}$,
 with $S_\alpha$ defined in Property~$\ref{Pro_31}.4$. 
\end{TT}\ \\[-6.5ex]
\end{quote}

The previous Theorem indicates that the time limit and the
continuous limit do not commute. In particular, the difference
between the discretized dynamics and the continuous one can be made
small by increasing $N$, while it becomes large beyond the time
scale $j\simeq \frac{1}{\gamma}\frac{\log N}{\log
\eta}$. This phenomenon is the same as in quantum
chaos and points to discretization of phase space
(in the traditional semi--classical treatment of quantum systems),
rather than to non--commutativity, as the source
of the so--called \enfasi{logarithmic
breaking time}. The constant $\gamma$ is a \enfasi{form factor}, which
reflects the fine 
structure of the dynamics: for instance, in the case of quantum cat
maps\cite{Ben03:1}, $\gamma=2$.
\begin{quote}
\begin{NNN}{}\ \\[-2.5ex]
\label{added_rem_lyap} 
The parameter
$\gamma>3$ in Theorem~\ref{propval}  may seem overestimated if
compared with the 
case of the quantum Cat Map, where 
$\gamma=2$. As we shall see (in particular in the next
Proposition~\ref{Lemma1}), the upper bound for $\gamma$ is dictated
by the discontinuities of the \co{S}awtooth \co{M}aps, and not by
commutativity. The corresponding exponent assumes the lower value
$\gamma>1$ in the case of discretized Cat Maps, that include \co{S}awtooth
\co{M}aps with integer $\alpha$. This result will be
presented in a forthcoming 
paper~\cite{Cap05:1}, in which we study the
breaking time $\tau_\text{B}\pt{N}$, here $\frac{1}{\gamma}\frac{\log N}{\log
\eta}$, relative to the chaotic or non--chaotic properties of the
dynamics.
In particular, in the hyperbolic regime,
the parameter $\log\eta$ of Theorem~\ref{propval} is replaced by the
Lyapunov exponent $\log\lambda$ whereas, in the elliptic regime, the
two limits $j,N\longrightarrow\infty$ do commute and in the parabolic
one, the breaking time is given by
$\tau_\text{B}\pt{N}=N^{\frac{1}{\gamma}}$. 
\end{NNN}
\end{quote}
The proof of Theorem~\ref{propval}
consists of several steps,
among which the most important is a property, satisfied by our choice
of \co{L}attice \co{S}tates, which we shall call
\enfasi{dynamical localization}.

We give a full proof that our choice of \co{L}attice \co{S}tates
satisfies such property, since it represents 
a natural request that should be fulfilled
by any consistent discretization/de--discretization
(quantization/de--quantization) scheme.
\begin{quote}
\begin{NNS}{}\ \\[-6.5ex]
\begin{Ventry}{2)} 
\label{nuovo_nuovissimo}  
\item[(1)] In analogy to the quantum case, 
Dynamical
localization is what one expects from a good choice of
states suited the study of the continuous limit: in fact, it
essentially amounts to asking that \co{LS} 
remain decently localized around the continuous trajectories
while evolving with the corresponding discrete evolution. As we shall
see this is the case only on logarithmic time--scales. 
Informally, when $N\to\infty$, the quantities
\begin{equation}
\notag
K_{j}(\bs{x},\bs{y}) := \< C_\nh(\bs{x}), W_{\alpha,N}^{j} \:C_\nh(\bs{y})\>
\end{equation}
should behave as if
$\nh|K_j(\bs{x},\bs{y})|^2\simeq\delta(S_\alpha^j\:\bs{x}-\bs{y})$:
this would make the discretization analogous to
the notion of \enfasi{regular
quantization} described in Section V of~\cite{Slo94:1}.
Actually, with our choice of \co{LS}, the quantity
$K_{j}(\bs{x},\bs{y})$ is a Kronecker delta.

\item[(2)] In quantum chaos, instead of seeking for the dynamical
localization, one can study the \enfasi{dynamical spreading} of
\co{C}oherent \co{S}tates. Consider for instance the classical
function $f$ over the 
phase space, its corresponding quantum observable
${\text{Op}}_{\hbar}\pt{f}$ and a \co{C}oherent \co{S}tate
$\coh{\hbar}{\bs{x}}$ centred at the point $\bs{x}$. The 
time needed for the quantum mechanical expectation 
$\< C_\hbar(\bs{x}), {\text{Op}}_{\hbar}\pt{f} \:C_\hbar(\bs{x})\>$
to converge to
the average of $f$ over a suitable invariant measure
can be explicitly analyzed. Recent
work~\cite{Bou04:1,Sch04:1} shows that also this time scales
logarithmically in $\hbar$, at least for the automorphisms on the
$2$--torus.

\item[(3)] The constraint $j\le C \log \nh$ is typical of \enfasi{hyperbolic}
behavior with Lyapunov exponent $\log\lambda$ and comes heuristically as
follows: the expansion of an 
initial small distance $\delta$ can be exponential until the distance
becomes the largest possible, namely $\delta
\lambda^{T_{\text{B}}}\simeq 1$. 
After discretization, the minimal distance gives $\delta=\frac{1}{N}$,
therefore one estimates $T_{\text{B}}\simeq\frac{\log N}{\log
\lambda}$, which is called \enfasi{breaking time} and sets the
time--scale over which continuous and discretized dynamics mimic each
other. 

\item[(4)] In quantum chaos, the semi--classical analysis leads to an
estimate of $T_{\text{B}}$ exactly as above; further, 
the logarithmic dependence on $\hbar$ of
$T_{\text{B}}$ is a signature of the hyperbolic character of the
classical limit. Conversely, if the classical limit is regular, then
the time scale when quantum and classical behaviors are more or less
indistinguishable goes as $\hbar^{-b},\ b>0$. 
Another interpretation of the breaking time is given in~\cite{Fnj04:1},
where it is related 
to the shortest time needed for the system to transfer all scales
$1\geq\ell\geq\hbar$ down to the ``quantum scale'' $\hbar$. Indeed,
this is the scale at which the differences among quantum and classical
mechanics come up.
Regarding the \co{SM},
the hyperbolic case corresponds to $S_\alpha$ with eigenvalue
$\lambda>1$, whereas the regular cases are
the \enfasi{elliptic} one (two complex eigenvalues) and the
\enfasi{parabolic} one (only one eigenvalue $=1$).

\item[(5)] The dynamical localization property has fruitfully been used in
several quantum contexts~\cite{Ben03:1}; however, to our knowledge,
this is the first 
instance, though not properly quantal, where dynamical localization is
fully exposed.
\end{Ventry}
\end{NNS}
\end{quote}

Before proceeding with the proof of Theorem~\ref{propval}, it is
important to notice that in its statement the Lyapunov 
exponent $\log \lambda$ does
not appear but $\log \eta$, instead; of course 
$\lambda$ and $\eta$ are related for $\lambda$ is eigenvalue
of $S_\alpha$, and $\eta$ of $\sqrt{S_\alpha^{\dagger}S_\alpha}$ (see Remark~\ref{added_rem_lyap}).

\begin{figure}[H]
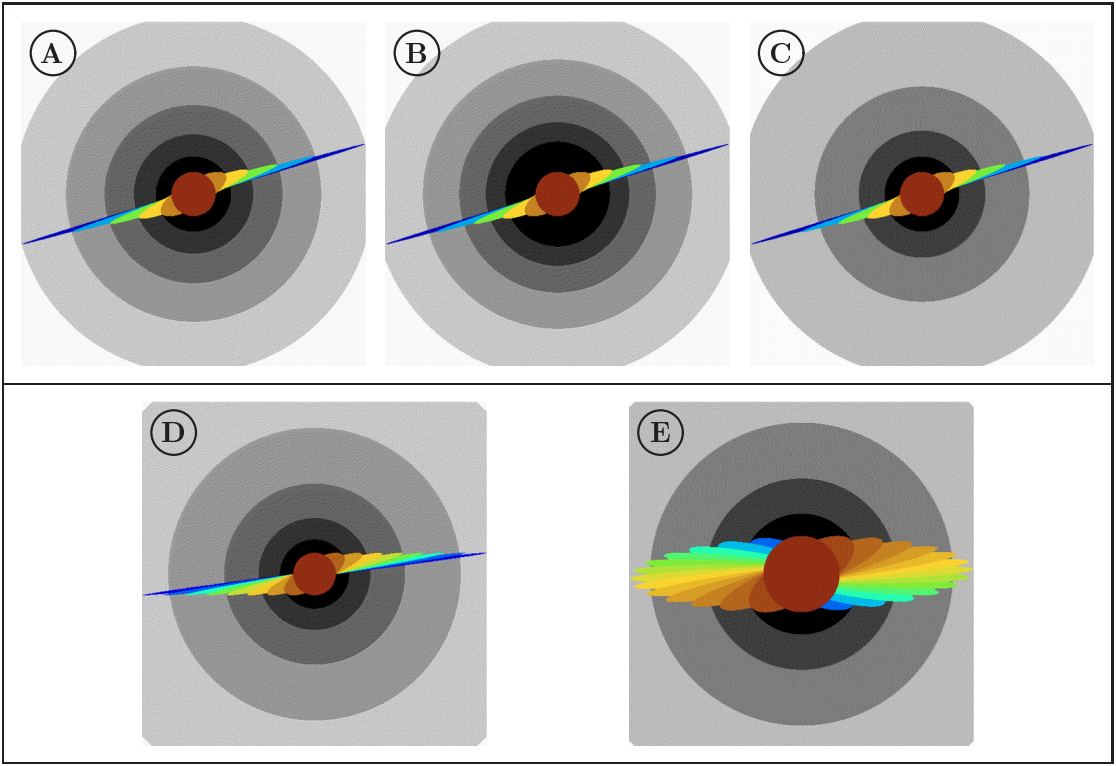

\begin{center}
\begin{picture}(248,170)(-4,-89) 

\put(-4,-89){\framebox(248,170){}}
\put(-4,-4){\framebox(248,85){}}
\thicklines
\put(0,0){\includegraphics[width=77\unitlength,height=77\unitlength]{./tru_005_541.epsi}}
\put(2,65){\makebox(10,10)[cc]{\bf A}}
\put(7,70){\circle{10}}
\put(81.5,0){\includegraphics[width=77\unitlength,height=77\unitlength]{./lya_005_541.epsi}}
\put(83.5,65){\makebox(10,10)[cc]{\bf B}}
\put(88.5,70){\circle{10}}
\put(163,0){\includegraphics[width=77\unitlength,height=77\unitlength]{./eta_005_541.epsi}}
\put(165,65){\makebox(10,10)[cc]{\bf C}}
\put(170,70){\circle{10}}
\put(27,-85){\includegraphics[width=77\unitlength,height=77\unitlength]{./eta_008_541.epsi}}
\put(29,-20){\makebox(10,10)[cc]{\bf D}}
\put(34,-15){\circle{10}}
\put(136,-85){\includegraphics[width=77\unitlength,height=77\unitlength]{./eta_015_541.epsi}}
\put(138,-20){\makebox(10,10)[cc]{\bf E}}
\put(143,-15){\circle{10}}

\thinlines

\end{picture}
\caption[]{In Plots {\bf A}, {\bf B} and {\bf C} we compare the
estimates of the (maximum) stretching given by the action of
the \co{SM} $S_{1/10}^{\phantom{n}}$ and its temporal iterates
$S_{1/10}^n$ ($n\leq 5$) given by $\lambda$, 
respectively $\eta$,  on a small ball $B_{\bs{v}}^0$ of radius
$\bs{v}$,  
centered in $\pt{\frac{1}{2},\frac{1}{2}}\in\IT$. The five
evolved images of the ball, namely $\pg{B_{\bs{v}}^n\ ,\ 1<n\leq 5}$,
are plotted together with $B_{\bs{v}}^0$,
using different colors. In {\bf A} we surround every evolved
ball $B_{\bs{v}}^n$ with the smallest circle containing it. We
compare that plot with {\bf B} and {\bf C}, in which the
surrounding circles have radii proportional to $\lambda^n \bs{v}$,
respectively $\eta^n \bs{v}$; 
in both cases the correct radii of {\bf A} are
overestimated although, on the long run, circles in {\bf B} provide a
good approximation.\\
The fake hyperbolicity given by $\eta$ is clearly shown in {\bf D} and
{\bf E}, where a parabolic \co{SM} $S_{0}$ and an elliptic one
$S_{-1/20}$ are presented: in the first case the maximum spreading
grows linearly, whereas in the second one it remains
confined, and the estimate given by the surrounding circles of radii
growing as powers of $\eta$ is inappropriate.\\
Note that in all examples {\bf C}--{\bf E}, the black
circles of radii $\eta \:\bs{v}$ rightly surround $B_{\bs{v}}^1$.}
\ 
\label{artefact}
\end{center}
\end{figure}

As will become clear during the proof, the use of $\eta$ and not of
$\lambda$ is required by the discontinuous character of
\co{SM}. In fact, the discontinuities do not allow us to control the
difference between the 
n--th iterates of the discretized and the continuous dynamics, but instead
force us to estimate that difference at each single time--step up to
$n$ and to put all
the estimates together. In the single time--step estimate,
independently of whether the map is continuous or not, one must use
$\eta$, which coincides with $\lambda$ only when the dynamical matrix
$S_\alpha$ is symmetric. Indeed, Figure~\ref{artefact}
shows that the eigenvalue $\eta$ correctly describes how 
volumes behaves under a single application of the dynamics, whereas
$\lambda$ underestimates it. On the contrary, it is $\lambda^n$ which
asymptotically controls the
stretching, whereas $\eta^n$ largely overestimates it. In the regular
elliptic case, where $\lambda=0$ and $\eta\geq\sqrt{2}$,  
the use of $\eta$ gives the impression of
hyperbolic stretching, whereas the elliptic motion is
confined: from the lower strip in Figure~\ref{artefact} it is apparent
that such hyperbolicity is spurious.
\ \\[-15.5ex]
\begin{quote}
\begin{TT}[Dynamical localization with $\bs{\{\vert
C_\nh(\bs{x})\rangle\}}$ states]{}\ \\[2ex]
\label{dynloc2}
\!\!For $\alpha\in\IR$,
$\beta\in\IR^+\setminus\left(0,2\:\right]$ and 
$d_0>0$, there exists $N_0=N_0(\alpha,\beta,d_0)\in\IN^+$
with the following property: if \mbox{$N > N_0$} and
$n<\frac{1}{\beta}\frac{\log N}{\log \eta}$,
then 
\begin{equation*}
d_{\IT}\pt{S_\alpha^n\pt{\bs{x}},\bs{y}} \geq d_0 \Longrightarrow \big<C_\nh(\bs{x})\,\big\vert\, W_{\alpha,N}^{n}\,C_\nh(\bs{y})\big>=
0\ ,
\end{equation*}
for all $\bs{y}\in\IT$ and
$\bs{x}\in{G_n^N}\pt{\frac{\widetilde{N}}{2N}}$, where
$W_{\alpha,N}^{n}$ is the unitary operator defined
in~\eqref{aggiunta},
$\widetilde{N}=2\,\sqrt{2}\pt{\sqrt{2}+1}\eta^{2n}$ and
${G_n^N}\pt{\varepsilon}$ has been introduced in Definitions~\ref{Gnbar}.
\end{TT}
\end{quote}
In order to prove Theorem~\ref{dynloc2}, we need the following 
result, whose proof can be found in Appendix~\ref{app_Lem_uno}.\\[-2ex]
\begin{quote}
\begin{PPP}{}\ \\[-1ex]
\label{Lemma2}
With the notation of Definitions~\ref{dont}
and~\ref{Gnbar}, and with $\pq{E}^{\circ}$ denoting the complement
of $E\subseteq\IT$, 
$\pq{E}^{\circ}\coleq\IT\setminus E$, the following inclusions
hold:
\begin{equation}
\pq{\overline{\Gamma}_n\pt{\varepsilon+ 
\frac{1}{\sqrt{2}N}}}^{\circ}\subseteq 
{G_n^N}\pt{\varepsilon}\subseteq \pq{\overline{\Gamma}_n\pt{\varepsilon-
\frac{1}{\sqrt{2}N}}}^{\circ}
\ \cdot
\label{lemma_1_5}
\end{equation}
Further, for $\alpha\in\IR$ and $n\in{\IN}^+$, if
\begin{multline}
N>\widetilde{N}=2\,\sqrt{2}\pt{\sqrt{2}+1}\eta^{2n}
\quad\text{and}\quad
\bs{x}\in {G_n^N}\pt{\frac{\widetilde{N}}{2N}}\quad
\text{then}\\
d_{\IT}\pt{\frac{U_\alpha^p\pt{N\bs{x}}}{N},
\frac{V_\alpha^p\pt{\hat{\bs{x}}_N}}{N}}\leq
\frac{\sqrt{2}}{N}\pt{\frac{\eta^{p+1}-1}{\eta-1}}\quad,\quad\forall p\leq
n\ \cdot
\label{lemma1_1}
\end{multline}
\end{PPP}\ \\[-7.5ex]
\end{quote}
\textbf{Proof of Theorem~\ref{dynloc2} :}\\[3ex]
Using the definition of $\{\vert
C_\nh(\bs{x})\rangle\}$ in~\eqref{CSforL1}, we easily compute
\begin{equation}
\big<C_\nh(\bs{x})\,\big\vert\, W_{\alpha,N}^{n}\,C_\nh(\bs{y})\big> =  
\Big\langle \hat{\bs{x}}_N\;\Big\vert\;
V_\alpha^{-n}
\pt{\hat{\bs{y}}_N}
\Big\rangle
=\delta^{(N)}_{\;
V_\alpha^n
\pt{\hat{\bs{x}}_N}
\:,\:\hat{\bs{y}}_N
}
\ \cdot
\label{loc_c31}
\end{equation}
Using the triangular inequality, we get:
\begin{multline}
d_{\IT}\pt{\frac{U_\alpha^n\pt{N\bs{x}}}{N}\,,\,\bs{y}}\leq
d_{\IT}\pt{\frac{U_\alpha^n\pt{N\bs{x}}}{N}\,,
\,\frac{V_\alpha^n\pt{\hat{\bs{x}}_N}}{N}} +\\
+ d_{\IT}\pt{\frac{V_\alpha^n\pt{\hat{\bs{x}}_N}}{N}\,,
\,\frac{\hat{\bs{y}}_N}{N}} +
d_{\IT}\pt{\frac{\hat{\bs{y}}_N}{N}\,,
\,\bs{y}}
\notag
\end{multline}
or equivalently, using the Definitions~\eqref{Ualpha},
\begin{multline}
d_{\IT}\pt{\frac{V_\alpha^n\pt{\hat{\bs{x}}_N}}{N}\,,
\,\frac{\hat{\bs{y}}_N}{N}}\geq 
d_{\IT}\pt{S_\alpha^n\pt{\bs{x}}\,,\,\bs{y}} -\\
-d_{\IT}\pt{\frac{U_\alpha^n\pt{N\bs{x}}}{N}\,,
\,\frac{V_\alpha^n\pt{\hat{\bs{x}}_N}}{N}} - 
d_{\IT}\pt{\frac{\hat{\bs{y}}_N}{N}\,,
\,\bs{y}}\ \cdot
\notag
\end{multline}
Now, since
$d_{\IT}\pt{S_\alpha^n\pt{\bs{x}}\,,\,\bs{y}}\geq d_0$ by hypothesis,
using~\eqref{nuovopt_2} in Appendix~\ref{app_Lem_uno} and observing 
that $\bs{x}\in{G_n^N}\pt{\frac{\widetilde{N}}{2N}}$
permits us to use~\eqref{lemma1_1} in Proposition~\ref{Lemma2}, 
namely that
\begin{equation}
N>\widetilde{N}\ \Longrightarrow\  
d_{\IT}\pt{\frac{U_\alpha^n\pt{N\bs{x}}}{N},  
\frac{V_\alpha^n\pt{\hat{\bs{x}}_N}}{N}}\leq
\frac{\sqrt{2}}{N}\pt{\frac{\eta^{n+1}-1}{\eta-1}}\ ,
\label{dYYN2}
\end{equation}
we can derive
\begin{equation}
d_{\IT}\pt{\frac{V_\alpha^n\pt{\hat{\bs{x}}_N}}{N}\,,
\,\frac{\hat{\bs{y}}_N}{N}}\geq 
d_0 - \frac{\sqrt{2}}{N}\pt{\frac{\eta^{n+1}-1}{\eta-1}} -
\frac{1}{\sqrt{2}N} \ \cdot
\notag
\end{equation}
The r.h.s. of the previous inequality can always be made strictly
larger than $\frac{1}{N}$, 
\begin{equation}
d_{\IT}\pt{\frac{V_\alpha^n\pt{\hat{\bs{x}}_N}}{N}\,, 
\,\frac{\hat{\bs{y}}_N}{N}}>\frac{1}{N}\ ,
\label{dYYN4}
\end{equation}
by choosing an $N$ larger than
\begin{equation}
N_{\text{\co{M}}}\pt{n} 
=\max\pg{\frac{1}{d_0}\pq{1 + \sqrt{2}\pt{\frac{\eta^{n+1}-1}{\eta-1}} +
\frac{1}{\sqrt{2}}}\ ,\
\widetilde{N}=2\,\sqrt{2}\pt{\sqrt{2}+1}\eta^{2n}}\ , 
\label{loc_c347}
\end{equation}
so that the condition on the l.h.s. of~\eqref{dYYN2} is also
satisfied.
From~\eqref{loc_c31} and~\eqref{dYYN4}, we have
\begin{equation}
N>N_{\text{\co{M}}}\pt{n}\quad\Longrightarrow\quad
\big<C_\nh(\bs{x})\,\big\vert\, W_{\alpha,N}^{n}\,C_\nh(\bs{y})\big> =
0 \ \cdot
\label{loc_c347a}
\end{equation}
Indeed, if the toral distance between two points $\pt{\bs{z},\bs{w}}$
exceeds $\frac{1}{N}$, then the corresponding grid points
$\pt{\hat{\bs{z}}_N,\hat{\bs{w}}_N}$ are different and then the
periodic Kronecker delta in~\eqref{loc_c31} vanishes.

\noindent Since the (non--decreasing) function $N_{\text{\co{M}}}$
in~\eqref{loc_c347} is eventually bounded by 
$\eta^{\beta n}$ ($\beta$ being strictly greater than two), we define
$\overline{n}$ as the time when $\displaystyle
N_{\text{\co{M}}}\pt{\overline{n}}=\eta^{\beta \overline{n}}\eqcol
N_0$,
and choose $N>N_0$,
$\bs{x}\in{G_n^N}\pt{\frac{\widetilde{N}}{2N}}$. Thus, if
$0<n<\overline{n}$, then
$N>N_0=N_{\text{\co{M}}}\pt{\overline{n}}>N_{\text{\co{M}}}\pt{n}$,
whereas if $\overline{n}\leq n<\frac{1}{\beta}\frac{\log N}{\log
\eta}$, then $N>\eta^{\beta n}>N_{\text{\co{M}}}\pt{n}$
and~\eqref{loc_c347a} holds for all $0< 
n<\frac{1}{\beta}\frac{\log N}{\log 
\eta}$.\hfill$\qed$ \\[3ex] 
In order to proceed with the proof of Theorem~\ref{propval}, we
need another auxiliary result which is proved in
Appendix~\ref{app_Lem_due}. 
\begin{quote}\ \\[-8ex]
\begin{PPP}{}\ \\[-1.5ex]
\label{Lemma1}
\noindent With the notation of Definition~\ref{Gnbar}, the following
relations hold for all
$p\in{\IN}$, $n\in{\IN}^{+}$ and 
$\varepsilon\in{\IR}^+$:
\begin{subequations} 
\label{lemma_1_0}
\begin{align} 
l\pt{\gamma_p}& \leq\eta^p
\ ,
\label{lemma_1_1}
\\\displaybreak
\mu\pt{\overline{\gamma}_p\pt{\varepsilon}}& \leq 2 \,\varepsilon\,
\eta^p + \pi\varepsilon^2 
\ ,
\label{lemma_1_2}
\\
\mu\pt{\overline{\Gamma}_n\pt{\varepsilon}} & \leq 
2\pt{\sqrt{2}+1}\varepsilon\,\eta^n + \pi\,n\,\varepsilon^2
\ \cdot
\label{lemma_1_3}
\end{align} 
Moreover, if $N\in\IN^{+}$ and
$\widetilde{N}=2\,\sqrt{2}\pt{\sqrt{2}+1}\eta^{2n}$
(cfr. equation~\eqref{lemma1_1} in Proposition~\ref{Lemma2}):
\begin{equation}
N>\widetilde{N}\Longrightarrow
\mu
\pt{{\pq{G_n^N \pt{\frac{\widetilde{N}}{2N}}}}^{\circ}}  
\leq 
\frac{38\,\eta^{3n}}{N}
\ \cdot
\label{lemma_1_4}
\end{equation}
\end{subequations}
\end{PPP}
\end{quote}
We are finally in position to conclude with\\[3ex]
\textbf{Proof of Theorem~\ref{propval}:}\\[3ex]
We subdivide the proof in two steps: in the first we concentrate on 
continuous $f$, that is $f\in\Ac\pt{\subset\Lspace{2}{\IT}}$; in the
second one we
extend the result to essentially bounded function by
applying the following Corollary of Lusin's
Theorem~\cite{Hew69:1,Rud87:1,Rie55:1}:\\[-6ex]
\begin{quote}
\it
Given $f\in\Lspace{\infty}{\cal X}$,
with $\c X$ compact,
there exists a sequence $\{f_n\}$ of continuous functions on $\c X$
such that $|f_n| \le \|f\|_\infty$ and converging to $f$ $\mu$ -- almost
everywhere. 
\end{quote}\ \\[-6ex]
\noindent $\bs{(1)}$ Let $f\in\Ac$ and $\displaystyle
{\text{Op}}_{j,N}\pt{f}\coleq\pt{\Theta_\alpha^j- {\cal 
J}_{\infty,\nh}\circ\Theta_ {\nh,\alpha}^j\circ {\cal
J}_{\nh,\infty}}\pt{f}$: notice that ${\text{Op}}_{j,N}\pt{f}$ is a
multiplication operator on $\Lspace{2}{\IT}$, but also an
$\Lspace{\infty}{\IT}$ \big(and thus also an $\Lspace{2}{\IT}$\big)
function. According to~\eqref{added2}, we must show that
 \begin{equation}
\forall g\in \Lspace{2}{\IT}\quad
 , \quad  
 \lim_{\substack{j,N\to\infty\\
j<\frac{1}{\gamma}\frac{\log N}{\log
\eta}}} 
\norm{\;{\text{Op}}_{j,N}\pt{f}\;g\;
}{2}
= 0\quad \cdot
\notag
 \end{equation}
Using Schwartz's inequality first with $g$ in the class of
\enfasi{simple functions} and then
using their density in $\Lspace{2}{\IT}$, we have just to show that
 \begin{equation}
\lim_{\substack{j,N\to\infty\\
j<\frac{1}{\gamma}\frac{\log N}{\log
\eta}}} 
\norm{\;{\text{Op}}_{j,N}\pt{f}\;}{2}
= 0\quad \cdot
\notag
 \end{equation}
Explicitly, using~\eqref{omegamu}, we write:
\begin{align}
\norm{\;{\text{Op}}_{j,N}\pt{f}\;}{2}^2 & =\om\pt{\,{\text{Op}}_{j,N}\pt{f}^*
{\text{Op}}_{j,N}\pt{f}\,} =
\om\pq{\pt{\Theta_\alpha^j f}^*\pt{\Theta_\alpha^j f}} +\nonumber\\
& \quad + \om\pq{\pt{{\cal J}_{\infty,\nh}\circ\Theta_ {\nh,\alpha}^j\circ
{\cal J}_{\nh,\infty}}\pt{f}^*\pt{{\cal J}_{\infty,\nh}\circ\Theta_
{\nh,\alpha}^j\circ {\cal J}_{\nh,\infty}}\pt{f}} +\nonumber\\
& \quad-2\;{\Re}\pg{\om\pq{\pt{\Theta_\alpha^j f}^*\pt{{\cal
J}_{\infty,\nh}\circ\Theta_ {\nh,\alpha}^j\circ {\cal
J}_{\nh,\infty}}\pt{f}}}\ ,\nonumber
\intertext{which, via Proposition~\ref{prop1}.1, becomes}
& = \om\pq{\Theta_\alpha^j \pt{\overline{f}} \Theta_\alpha^j \pt{f}} -
2\;{\Re}\pg{\tn\pq{{\cal J}_{\nh,\infty}\pt{\Theta_\alpha^j
f}^*\pt{\Theta_ {\nh,\alpha}^j\circ {\cal J}_{\nh,\infty}}\pt{f}}} +
\nonumber\\ 
& \quad + \tn\pq{\pt{{\cal J}_{\nh,\infty}\circ{\cal J}_{\infty,\nh}\circ\Theta_ {\nh,\alpha}^j\circ
{\cal J}_{\nh,\infty}}\pt{f}^*\pt{\Theta_
{\nh,\alpha}^j\circ {\cal J}_{\nh,\infty}}\pt{f}}\nonumber\ ,
\intertext{that, using Proposition~\ref{prop1}.3, can be recast as}
& = \pt{\om\circ\Theta_\alpha^j }\pt{\overline{f} f}  
+ \tn\pq{\pt{\Theta_ {\nh,\alpha}^j\circ
{\cal J}_{\nh,\infty}}\pt{f}^*\pt{\Theta_
{\nh,\alpha}^j\circ {\cal J}_{\nh,\infty}}\pt{f}} +
\nonumber\\ 
& \quad 
- 2\;{\Re}\pg{\tn\pq{\Big({\cal J}_{\nh,\infty}\circ\Theta_\alpha^j\Big)
\pt{f}^*\pt{\Theta_ {\nh,\alpha}^j\circ {\cal J}_{\nh,\infty}}\pt{f}}} 
\nonumber\\
& = \om\pt{{\abs{f}}^2} + \pt{\tn\circ\Theta_ {\nh,\alpha}^j}\pq{{\cal
J}_{\nh,\infty}\pt{f}^* {\cal J}_{\nh,\infty}\pt{f}} -
2\;{\Re}\pt{I_{j,N}\pt{f}} \ ,\nonumber
\notag
\intertext{with}
I_{j,N}\pt{f}& \coleq \tn\pq{\Big({\cal J}_{\nh,\infty}\circ\Theta_\alpha^j\Big)
\pt{f}^*\pt{\Theta_ {\nh,\alpha}^j\circ {\cal
J}_{\nh,\infty}}\pt{f}}\nonumber\\
& = \displaystyle  \nh \,\int_{\IT} \mu(\ud\bs{x})\, \int_{\IT}
 \mu(\ud\bs{y})\,  \overline{f(\bs{y})}\, f(S_\alpha^j \bs{x})  |\<
  C_\nh(\bs{x}), W_{\alpha,N}^j C_\nh(\bs{y}) \>
 |^2 \ \cdot\nonumber
\end{align}
Now, Proposition~\ref{prop1}.2 yields
\begin{equation}
\pt{\tn\circ\Theta_ {\nh,\alpha}^j}\pq{{\cal
J}_{\nh,\infty}\pt{f}^* {\cal J}_{\nh,\infty}\pt{f}}= \tn\pq{{\cal
J}_{\nh,\infty}\pt{f}^* {\cal
J}_{\nh,\infty}\pt{f}}\xrightarrow[N\longrightarrow\infty]{}
\om\pt{{\abs{f}}^2} \nonumber\ ,
\end{equation}
so that the strategy is to prove that also $I_{j,N}\pt{f}$
goes to $\displaystyle\om\pt{{\abs{f}}^2}=\int_{\IT} \mu(\ud\bs{x})
|f(\bs{x})|^2$ when $j,N\to\infty$ with $j<\frac{1}{\gamma}\frac{\log
N}{\log\eta}$.  

Resorting to
${G_n^N}\pt{\frac{\widetilde{N}}{2N}}$ in 
Definition~\ref{Gnbar}, and to its complementary set\\
$\pq{G_n^N\pt{\frac{\widetilde{N}}{2N}}}^{\circ} =
{\IT}\setminus{G_n^N}\pt{\frac{\widetilde{N}}{2N}}$, we can write
\begin{align}
 &\left| I_{j,N}\pt{f} -
 \int_{\IT} \mu(\ud\bs{y})\, |f(\bs{y})|^2 \right| \nonumber\\
 &=  \left| \int_{\IT} \mu(\ud\bs{x})\, \int_{\IT} \mu(\ud\bs{y})\,
 \overline{f(\bs{y})}\, 
 \bigl( f(S_\alpha^j \bs{x}) - f(\bs{y}) \bigr)\, \nh|\< C_\nh(\bs{x}), W_{\alpha,N}^{j} C_\nh(\bs{y})\>|^2
 \right| \nonumber\\
 & \le  \left| \int_{\pq{G_n^N\pt{\frac{\widetilde{N}}{2N}}}^{\circ}} \mu(\ud\bs{x})\, \int_{\IT} \mu(\ud\bs{y})\,
 \overline{f(\bs{y})} \bigl(f(S_\alpha^j \bs{x}) - f(\bs{y})\bigr) \nh|\< C_\nh(\bs{x}),
 W_{\alpha,N}^{j} C_\nh(\bs{y})\>|^2 \right| \nonumber\\
 &+ \left| \int_{{G_n^N}\pt{\frac{\widetilde{N}}{2N}}} \mu(\ud\bs{x})
 \int_{\IT} \mu(\ud\bs{y})
 \overline{f(\bs{y})}\bigl(f(S_\alpha^j \bs{x}) - f(\bs{y})\bigr) \nh|\< C_\nh(\bs{x}),
 W_{\alpha,N}^{j} C_\nh(\bs{y})\> |^2 \right|.
\label{secondterm}
\intertext{For the first integral in the r.h.s. of the previous
 expression we have:}
& \left| \int_{\pq{G_n^N\pt{\frac{\widetilde{N}}{2N}}}^{\circ}} \mu(\ud\bs{x})\, \int_{\IT} \mu(\ud\bs{y})\,
 \overline{f(\bs{y})} \bigl(f(S_\alpha^j \bs{x}) - f(\bs{y})\bigr) \nh|\< C_\nh(\bs{x}),
 W_{\alpha,N}^{j} C_\nh(\bs{y})\>|^2 \right| \nonumber\\
& \le 2 {\pt{\norm{f}{\infty}}}^2
\int_{\pq{G_n^N\pt{\frac{\widetilde{N}}{2N}}}^{\circ}}\mu(\ud\bs{x})\, \int_{\IT} \mu(\ud\bs{y})\, 
  \nh|\< {\pt{W_{\alpha,N}^{*}}}^j C_\nh(\bs{x}),
  C_\nh(\bs{y})\>|^2  \nonumber\\
& \le 2 {\pt{\norm{f}{\infty}}}^2
\mu\pt{\pq{G_n^N\pt{\frac{\widetilde{N}}{2N}}}^{\circ}}\leq
\frac{76\,\eta^{3j}}{N}{\pt{\norm{f}{\infty}}}^2\nonumber
\end{align}
where we have used \co{completeness} and \co{normalization} Properties~\ref{coh} and equation~\eqref{lemma_1_4} from
Proposition~\ref{Lemma1}; this term becomes negligible for
large $N>\widetilde{N}$ iff $j<\frac{1}{\gamma}\frac{\log N}{\log
\eta}$, with $\gamma>3$.

Now it remains to prove that the second term in~\eqref{secondterm} is
also negligible for 
large $N$:
 selecting a ball $B(S_\alpha^j \bs{x},d_0)$,
one derives 
\begin{align*}
 &\left| \int_{{G_n^N}\pt{\frac{\widetilde{N}}{2N}}} \mu(\ud\bs{x})\, \int_{\IT} \mu(\ud\bs{y})\, \overline{f(\bs{y})}\,
 \bigl( f(S_\alpha^j \bs{x}) - f(\bs{y}) \bigr)\, \nh|\< C_\nh(\bs{x}), W_{\alpha,N}^{j} C_\nh(\bs{y})\>|^2
 \right| \\
 & \le  \left| \int_{{G_n^N}\pt{\frac{\widetilde{N}}{2N}}} \mu(\ud\bs{x})\,
 \int_{B(S_\alpha^j \bs{x},d_0)} \mu(\ud\bs{y})\,
 \overline{f(\bs{y})} \bigl(f(S_\alpha^j \bs{x}) - f(\bs{y})\bigr) \nh|\< C_\nh(\bs{x}),
 W_{\alpha,N}^{j} C_\nh(\bs{y})\>|^2 \right| \\
 &+ \left| \int_{{G_n^N}\pt{\frac{\widetilde{N}}{2N}}} \mu(\ud\bs{x})
 \int_{{\IT}\setminus B(S_\alpha^j \bs{x},d_0)} \mu(\ud\bs{y})
 \overline{f(\bs{y})}\bigl(f(S_\alpha^j \bs{x}) - f(\bs{y})\bigr) \nh|\< C_\nh(\bs{x}),
 W_{\alpha,N}^{j} C_\nh(\bs{y})\> |^2 \right|.
\end{align*}
Applying the mean value theorem in the first double integral, we get that 
$\exists \:\bs{c} \in B(S_\alpha^j \bs{x}, d_0)$ such that
\begin{align*}
 &\left| \int_{{G_n^N}\pt{\frac{\widetilde{N}}{2N}}} \mu(\ud\bs{x})\, \int_{\IT} \mu(\ud\bs{y})\, \overline{f(\bs{y})}\,
 \bigl( f(S_\alpha^j \bs{x}) - f(\bs{y}) \bigr)\, \nh|\< C_\nh(\bs{x}), W_{\alpha,N}^{j} C_\nh(\bs{y})\>|^2
 \right| \\
 &\le  \int_{{G_n^N}\pt{\frac{\widetilde{N}}{2N}}} \mu(\ud\bs{x})\,\left| 
 \overline{f(\bs{c})}\, \bigl(f(S_\alpha^j \bs{x}) -
 f(\bs{c})\bigr)\right|\,
 \int_{B(S_\alpha^j
 \bs{x},d_0)} \mu(\ud\bs{y})\,\nh|\< {\pt{W_{\alpha,N}^{*}}}^j C_\nh(\bs{x}),  C_\nh(\bs{y})\>|^2  \\
 &\quad+ 2 \|f\|_{\infty}^{\:2}\int_{{G_n^N}\pt{\frac{\widetilde{N}}{2N}}} \mu(\ud\bs{x})
 \int_{{\IT}\setminus B(S_\alpha^j \bs{x},d_0)} \mu(\ud\bs{y})
 \nh|\< C_\nh(\bs{x}),
 W_{\alpha,N}^{j} C_\nh(\bs{y})\> |^2 \ \cdot
\intertext{Finally, using \co{completeness} and \co{normalization} 
(Properties~\ref{coh}), we arrive at the upper bound}
 &\le  
\;\|f\|_{\infty}\sup_{\substack{\bs{z}\in\IT\\
\bs{c}\in B(\bs{z},d_0)}}
\left|
 \bigl(f(\bs{z}) -
 f(\bs{c})\bigr)\right| 
+ 2 \;\|f\|_{\infty}^{\:2}\quad\nh
\sup_{\substack{\bs{x}\in{G_n^N}\pt{\frac{\widetilde{N}}{2N}}\\
\bs{y}\not\in B(S_\alpha^j \bs{x},d_0)}}
  |\< C_\nh(\bs{x}),
 W_{\alpha,N}^{j} C_\nh(\bs{y})\> |^2 \ \cdot
\end{align*}
By uniform continuity, the first term can be made arbitrarily
small, provided  we choose $d_0$ small enough. For
the second integral, we use Theorem~\ref{dynloc2},
which provides us with $N_0=N_0(d_0)$ depending on the same $d_0$
, such that the second term vanishes
for all $N>N_0$ and far all $j<\frac{1}{\gamma}\frac{\log N}{\log
\eta}$.

\noindent $\bs{(2)}$
In order to extend the result of point $\bs{(1)}$
to $f\in\Lspace{\infty}{\IT}$,
we use the Corollary of Lusin's Theorem, choose a
sequence ${\pg{f_n}}_n$ as in its statement and estimate
\begin{equation}
\lim_{\substack{j,N\to\infty\\
j<\frac{1}{\gamma}\frac{\log N}{\log
\eta}}} 
\norm{\;{\text{Op}}_{j,N}\pt{f}\;}{2}\leq
\lim_{\substack{j,N\to\infty\\
j<\frac{1}{\gamma}\frac{\log N}{\log
\eta}}} 
\norm{\;{\text{Op}}_{j,N}\pt{f-f_n}\;}{2}+
\lim_{\substack{j,N\to\infty\\
j<\frac{1}{\gamma}\frac{\log N}{\log
\eta}}} 
\norm{\;{\text{Op}}_{j,N}\pt{f_n}\;}{2}\nonumber\ \cdot
\end{equation}

Using point $\bs{(1)}$, the second term in the r.h.s. of
the previous equation can be bounded by arbitrarily small $\varepsilon$,
indeed $f_n\in\Ac$.

For the first term we proceed as follows: using
Definition~\ref{ThetaNSa} together with equations~\eqref{AWJNI1}
and~\eqref{AWJIN1due} of Appendix~\ref{AWPSCS}, we find 
\begin{equation}
\pt{{\cal J}_{\infty,\nh}\circ\Theta_ {\nh,\alpha}^j\circ{\cal
J}_{\nh,\infty}}(g)(\bs{x}) 
=\sum_{\bs{\ell}\in\ZNZD} 
\Gamma_N\pt{g}\pt{\frac{V_\alpha\pt{\bs{\ell}}}{N}}
\;\car_{Q_N\pt{\frac{\bs{\ell}}{N}}}(\bs{x})
\label{JTJauno}
\ ,
\end{equation}
where $g$ is any measurable function on $\IT$.
Then, because of how the running average operator (\co{RAO}) $\Gamma_N$ is defined, 
for all $g\in \Lspace{1}{\IT}$ it follows that
\begin{align}
\norm{\pt{{\cal J}_{\infty,\nh}\circ\Theta_ {\nh,\alpha}^j\circ{\cal
J}_{\nh,\infty}}(g)}{1} & \leq
\norm{\pt{{\cal J}_{\infty,\nh}\circ\Theta_ {\nh,\alpha}^j\circ{\cal
J}_{\nh,\infty}}(\abs{g})}{1} 
=\norm{g}{1}
\nonumber\ ,
\intertext{where $\norm{\cdot}{1}$ denotes the
$\Lspace{1}{\IT}$--norm, and that} 
\norm{\pt{{\cal J}_{\infty,\nh}\circ\Theta_ {\nh,\alpha}^j\circ{\cal
J}_{\nh,\infty}}(g)}{\infty} 
& = \sup_{\bs{\ell}\in\ZNZD}\pg{\abs{\Gamma_N\pt{g}\pt{\frac{\bs{\ell}}{N}}}}
\leq\norm{\Gamma_N\pt{g}}{0}\leq\norm{g}{\infty}
\nonumber\ \cdot
\end{align}
\noindent
Indeed, the first equality in the last formula comes from the
definition of essential norm~\cite{Hew69:1} (which in this
case amounts to 
the greater absolute value assumed by the simple function ${\cal
J}_{\infty,\nh}\circ\Theta_ {\nh,\alpha}^j\circ{\cal
J}_{\nh,\infty}$), whereas the first inequality is a consequence of the
continuity of $\Gamma_N$ and the last one from
Proposition~\ref{runaveprps}.
Putting last two inequality together, we obtain 
\begin{equation}
\norm{\pt{{\cal J}_{\infty,\nh}\circ\Theta_ {\nh,\alpha}^j\circ{\cal
J}_{\nh,\infty}}(g)}{2} \leq \norm{g}{\infty}\norm{g}{1}
\nonumber\ ,
\end{equation}
whence, setting $g=f-f_n$,
\begin{align}
\norm{\;{\text{Op}}_{j,N}\pt{f-f_n}\;}{2}& =
\norm{\;\Theta_\alpha^j\pt{f-f_n}- {\cal 
J}_{\infty,\nh}\circ\Theta_ {\nh,\alpha}^j\circ {\cal
J}_{\nh,\infty}\pt{f-f_n}\;}{2}\\
& \leq
\norm{\;f-f_n\;}{2} + \norm{\;f-f_n\;}{\infty}\norm{\;f-f_n\;}{1}
\nonumber\quad ,\quad\forall j,N\ \cdot
\end{align}
Now convergence follows from Lusin's Corollary.\hfill$\qed$
\newpage
\begin{figure}[H]
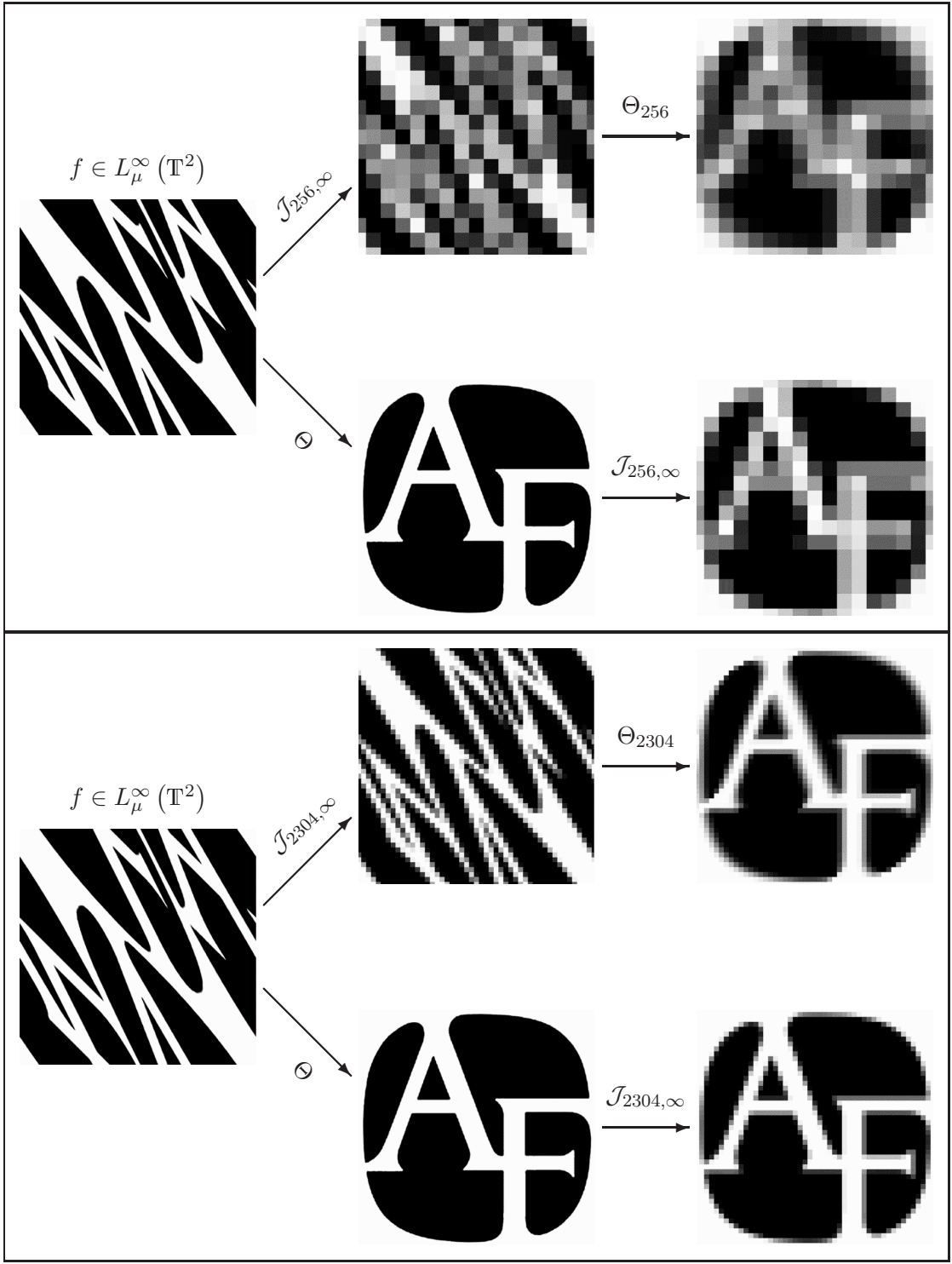

\begin{center}
\begin{picture}(248,165)(-4,-4) 

\put(-4,-4){\framebox(248,165){}}
\put(0,47.5){\includegraphics[width=62\unitlength,height=62\unitlength]{./o___001_016.epsi}}
\put(89,0){\includegraphics[width=62\unitlength,height=62\unitlength]{./ec__001_016.epsi}}
\put(89,95){\includegraphics[width=62\unitlength,height=62\unitlength]{./d___016_016.epsi}}
\put(178,0){\includegraphics[width=62\unitlength,height=62\unitlength]{./ecd_001_016.epsi}}
\put(178,95){\includegraphics[width=62\unitlength,height=62\unitlength]{./ded_001_016.epsi}}

\thicklines

\put(64,67.5){\vector(1,-1){23}}
\put(64,89.5){\vector(1,1){23}}
\put(153,31){\vector(1,0){23}}
\put(153,126){\vector(1,0){23}}

\put(0,109.5){\makebox(62,15.556349186)[cc]{$f\in\Lspace{\infty}{\IT}$}}
\put(62,47.5){\begin{rotate}{-45}\makebox(19.798989873,15.556349186)[cc]{$\Theta$}\end{rotate}}
\put(73,98.5){\begin{rotate}{45}\makebox(19.798989873,15.556349186)[cc]{${\cal J}_{256,\infty}$}\end{rotate}}
\put(153,31){\makebox(23,15.556349186)[cc]{${\cal J}_{256,\infty}$}} 
\put(153,126){\makebox(23,15.556349186)[cc]{$\Theta_{256}$}}

\thinlines

\end{picture}
\label{uno}
\begin{picture}(248,165)(-4,-4) 

\put(-4,-4){\framebox(248,165){}}

\put(0,47.5){\includegraphics[width=62\unitlength,height=62\unitlength]{./o___001_016.epsi}}
\put(89,0){\includegraphics[width=62\unitlength,height=62\unitlength]{./ec__001_016.epsi}}
\put(89,95){\includegraphics[width=62\unitlength,height=62\unitlength]{./d___048_048.epsi}}
\put(178,0){\includegraphics[width=62\unitlength,height=62\unitlength]{./ecd_001_048.epsi}}
\put(178,95){\includegraphics[width=62\unitlength,height=62\unitlength]{./ded_001_048.epsi}}

\thicklines

\put(64,67.5){\vector(1,-1){23}}
\put(64,89.5){\vector(1,1){23}}
\put(153,31){\vector(1,0){23}}
\put(153,126){\vector(1,0){23}}

\put(0,109.5){\makebox(62,15.556349186)[cc]{$f\in\Lspace{\infty}{\IT}$}}
\put(62,47.5){\begin{rotate}{-45}\makebox(19.798989873,15.556349186)[cc]{$\Theta$}\end{rotate}}
\put(73,98.5){\begin{rotate}{45}\makebox(19.798989873,15.556349186)[cc]{${\cal J}_{2304,\infty}$}\end{rotate}}
\put(153,31){\makebox(23,15.556349186)[cc]{${\cal J}_{2304,\infty}$}} 
\put(153,126){\makebox(23,15.556349186)[cc]{$\Theta_{2304}$}} 

\thinlines

\end{picture}
\caption[]{These two plots show how the difference between
${\cal J}_{\nh,\infty}\circ\Theta_\alpha$ and 
$\Theta_{\nh,\alpha}\circ{\cal J}_{\nh,\infty}$ becomes smaller with 
$N$. For the continuous \co{SM},
$\Theta_1$, the actions ${\cal J}_{\nh,\infty}\circ\Theta_1$ and 
$\Theta_{\nh,1}\circ{\cal J}_{\nh,\infty}$
on $f\in\Lspace{\infty}{\IT}$ (left part of both
plots) are plotted for two different $N$: $N=16$ (top) and $N=48$ (bottom).
The resulting matrices are mapped back,
together with the function $\Theta_1\pt{f}$, on
the unfolded torus, by means of the de--discretization operator 
${\cal J}_{\infty,\nh}$.} 
\ 
\label{due}
\end{center}
\end{figure}
\newpage

\clearpage

\clearpage
\begin{figure}[H]
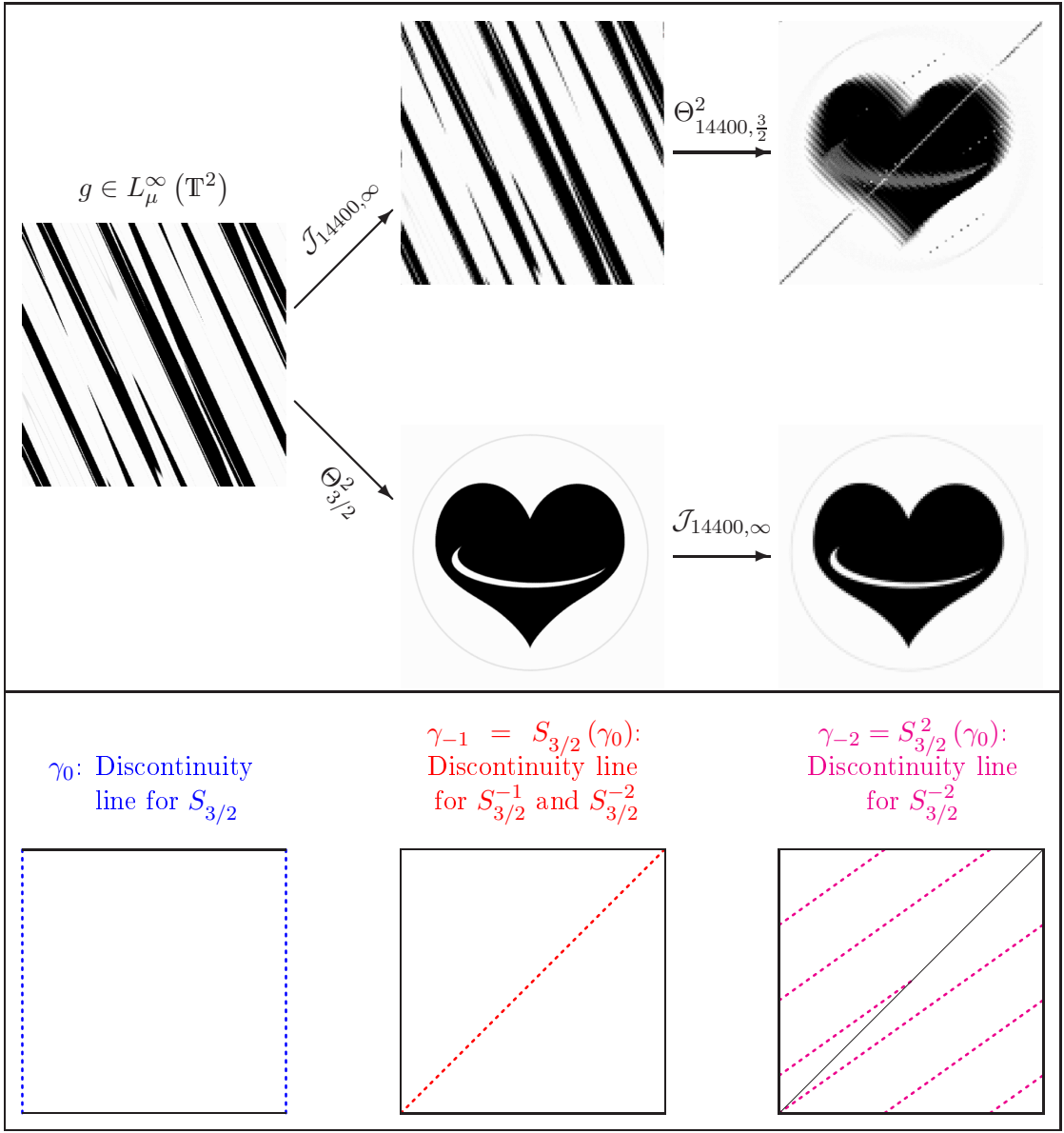

\begin{center}
\begin{picture}(248,265)(-4,-4) 

\put(-4,-4){\framebox(248,265){}}
\put(-4,-4){\framebox(248,103){}}

\put(0,147.5){\includegraphics[width=62\unitlength,height=62\unitlength]{./o___001_sm2.epsi}}
\put(89,100){\includegraphics[width=62\unitlength,height=62\unitlength]{./ec__001_sm2.epsi}}
\put(89,195){\includegraphics[width=62\unitlength,height=62\unitlength]{./d___120_sm2.epsi}}
\put(178,100){\includegraphics[width=62\unitlength,height=62\unitlength]{./ecd_001_sm2.epsi}}
\put(178,195){\includegraphics[width=62\unitlength,height=62\unitlength]{./ded_001_sm2.epsi}}

\thicklines

\put(64,167.5){\vector(1,-1){23}}
\put(64,189.5){\vector(1,1){23}}
\put(153,131){\vector(1,0){23}}
\put(153,226){\vector(1,0){23}}
\ccbl
\dottedline{2}(0,0)(0,62)
\dottedline{2}(62,0)(62,62)
\put(4,87.5){
\parbox[t]{30mm}{\ \\$\gamma_{0}$: Discontinuity\\ 
\phantom{$\gamma_{0}$: }line for $S_{3/2}^{\phantom{-1}}$}} 
\ccro
\put(93,87.5){	\parbox[t]{30mm}{$\gamma_{-1}
		=S_{3/2}^{\phantom{-1}}\pt{\gamma_{0}}$:  
 		Discontinuity line\\\phantom{i}for $S_{3/2}^{-1}$ and $S_{3/2}^{-2}$
		}}
\dottedline{2}(89,0)(151,62)
\ccvi
\put(182,87.5){	 \parbox[t]{30mm}{\begin{center}$\gamma_{-2}
		=S_{3/2}^{\:2}\pt{\gamma_{0}}$:  
 		Discontinuity line for $S_{3/2}^{-2}$
		\end{center}}}
\dottedline{2}(178,0)(240,44.285714)
\dottedline{2}(178,44.285714)(202.8,62)
\dottedline{2}(202.8,0)(240,26.571428)
\dottedline{2}(178,26.571428)(227.6,62)
\dottedline{2}(227.6,0)(240,8.857142)
\dottedline{2}(178,8.857142)(209,31)
\nncc
\thinlines

\put(0,209.5){\makebox(62,15.556349186)[cc]{$g\in\Lspace{\infty}{\IT}$}}
\put(62,147.5){\begin{rotate}{-45}\makebox(19.798989873,15.556349186)[cc]{$\Theta_{3/2}^2$}\end{rotate}}
\put(73,198.5){\begin{rotate}{45}\makebox(19.798989873,15.556349186)[cc]{${\cal J}_{14400,\infty}$}\end{rotate}}
\put(153,131){\makebox(23,15.556349186)[cc]{${\cal J}_{14400,\infty}$}} 
\put(153,226){\makebox(23,15.556349186)[cc]{$\Theta_{14400,\frac{3}{2}}^2$}}

\put(0,0){\line(1,0){62}}
\put(0,62){\line(1,0){62}}
\put(178,0){\line(1,1){62}}
\put(89,0){\framebox(62,62){}}
\put(178,0){\framebox(62,62){}}

\end{picture}
\caption[]{Here, the same picture as in Figure~\ref{due}, is represented, with
a finer discretization given by $N=120$ and a different function 
$g\in\Lspace{\infty}{\IT}$, 
for a \underline{discontinuous} \co{SM}, $\Theta_{3/2}$, acting two times. 
Choosing a function $g$ 
with
sharp variation across $\gamma_{0}$ (blue lines), the
preimage of $\gamma_{-1}$, the discontinuity of $\Theta_{3/2}$ makes
it evident how the 
differences between 
${\cal J}_{14400,\infty}\circ\Theta_\alpha^2$ and 
$\Theta_{14400,\alpha}^2\circ{\cal J}_{14400,\infty}$
are the greater the closer they are to the
discontinuity line $\gamma_{-1}$ (red lines).  Of course, the longer
the temporal evolution, the worst the correspondence, in the sense
that several new 
discontinuity lines come to play a role. In the case at hands,
the map acts twice, and $\gamma_{-2}$ is felt
by
$\Theta_{14400,\alpha}^2\circ{\cal J}_{14400,\infty}$, as expected.}
\ 
\label{tre}
\end{center}
\end{figure}
\section{Conclusions}
\label{concl}
\vspace{3mm}
In this article we have considered discrete approximants of
\co{S}awtooth \co{M}aps on the torus and we have studied them 
in an algebraic framework
modeled on the so--called \co{A}nti--\co{W}ick quantization;
In fact, finite--dimensional discretization and
quantization can be seen as 
similar procedures in that they map an abelian Von~Neumann algebra (of
essentially bounded functions on phase--space) into
finite--dimensional matrix subalgebras, the only difference being
whether the latter are diagonal (commutative) or not.

In the semi--classical analysis of classically chaotic quantum
systems, the correspondence classical/quantum is
usually observed only on time--scales that are logarithmic in the
quantization parameter $\hbar$.
The motivation of our study was to show that the same phenomenon
arises when a hyperbolic classical system is discretized, namely forced to
move on a lattice, and afterwards the lattice spacing is sent to zero.

Previous results~\cite{Ben04:1} based on the numerical investigation of the
entropy production, indicate that it should indeed be so; however,
these results were not supported by a solid framework where to analyze
the continuous limit of the family of discrete approximants. This is
the content of this article.

The major difficulty was represented by the need of controlling the
discontinuous character of \co{S}awtooth \co{M}aps, which was made
possible by an appropriate choice of \co{L}attice \co{S}tates. In
fact, similarly to the entropic approach which, despite the dynamics
being classical, was based on a quantum
dynamical entropy, the discretization/de--discretization procedure we
set up is based on quantum tools. 

The choice of \co{L}attice \co{S}tates was naturally pointed to by
the lattice structure of the discrete phase--space and turned out to
posses the right \co{localization} properties for mastering the
discontinuities. The result is the appearance of a logarithmic
time--scale when the discrete hyperbolic \co{SM} tend to their
continuous limit; namely, the continuous and discrete dynamics agree
up to a \enfasi{breaking time} which is proportional to the logarithm
of the lattice spacing.

The proportionality constant does not involve the Lyapunov exponent,
that is the eigenvalue $\lambda>1$ of the dynamical matrix $S_\alpha$,
rather the largest eigenvalue, $\eta$, of
$\sqrt{S_\alpha^{\dagger}S_\alpha}$. In the case of elliptic \co{SM},
$\abs{\lambda}=1$, $\eta>\sqrt{2}$; however the resulting breaking
time is a spurious effect, while when $\lambda>1$, the presence of $\eta$
in the breaking time seems to be an unavoidable consequence of the
discontinuous dynamics.
\appendix
\section{\co{A}nti \co{W}ick discretization of
$\bs{\Lspace{\infty}{\IT}}$}
\label{AWPSCS}
\vspace{3mm}    
In this appendix we will apply Definitions~\ref{qWick} and
discretize $\Lspace{\infty}{\IT}$ by means of the \co{LS} set 
$\{\vert C_\nh(\bs{x})\rangle
\mid \bs{x}\in\IT\}\in {\cal H}_\nh$ introduced in Section~\ref{PSCS}.

\noindent In this framework, the discretizing/de--discretizing
operators of Definitions~\ref{qWick} read:
 \begin{align}
 {\Lspace{\infty}{\IT}\ni} f & \mapsto   
 N^2 \int_{\IT}\mu(\ud \bs{x})\, f(\bs{x})\,
 \vert \,\hat{\bs{x}}_N\,\rangle\langle \,\hat{\bs{x}}_N\,\vert
=:{\cal J}_{\nh,\infty}(f)\in \cal D_\nh\ ,
\label{AWSM_1}
\\
  \cal D_\nh \ni X & \mapsto 
  \langle \,\hat{\bs{x}}_N\,\vert\,X\,\vert\,\hat{\bs{x}}_N\,\rangle
  =:{\cal J}_{\infty,\nh}(X)(\bs{x})
 \in{\cal S}\pt{\IT}\subset
\Lspace{\infty}{\IT}
  \ ,
\label{AWSM_2}
 \end{align}
where ${\cal S}\pt{\IT}$ denotes the set of {\it simple
functions}~\cite{Hew69:1} 
on the torus. 
The matrix elements of ${\cal J}_{\nh,\infty}(f)$ are as follows:
\begin{align}
M_{\bs{\ell},\bs{m}}^{(f)}&\coleq
\bkkk{\,\bs{\ell}\,}{\,{\cal J}_{\nh,\infty}(f)\,}{\,
\bs{m}\,}
\notag
\\ 
&=N^2 \int_{\IT}\mu(\ud\bs{x})\ f(\bs{x})\
\langle\,\bs{\ell}\,
\vert \,\hat{\bs{x}}_N\,\rangle\langle \,\hat{\bs{x}}_N\,\vert
\,\bs{m}\,\rangle\notag\\
\notag
\\
&= 
N^2 
\int_{0}^{1}\ud x_1\,
\int_{0}^{1}\ud x_2\ f(\bs{x})\ 
\delta^{(N)}_{\ell_1\:,\:\hat{x}_{N,1}}\;
\delta^{(N)}_{\ell_2\:,\:\hat{x}_{N,2}}\;
\delta^{(N)}_{m_1\:,\:\hat{x}_{N,1}}\;
\delta^{(N)}_{m_2\:,\:\hat{x}_{N,2}}\;
\notag\\
\notag
\\
&= 
N^2 \ 
\delta^{(N)}_{\ell_1\:,\:m_1}\;
\delta^{(N)}_{\ell_2\:,\:m_2}\;
\int_{0}^{1}\ud x_1\,
\int_{0}^{1}\ud x_2\ f(\bs{x})\ 
\delta^{(N)}_{\ell_1\:,\:\floor{N x_1+\pum}}\;
\delta^{(N)}_{\ell_2\:,\:\floor{N x_2+\pum}}\;\ \cdot
\notag
\end{align}
This implies
\begin{equation}
M_{\bs{\ell},\bs{m}}^{(f)}
 = N^2 \;\delta^{(N)}_{\bs{\ell},\bs{m}}
\;
\int_{\frac{\ell_1-\pum}{N}}^{\frac{\ell_1+\pum}{N}}\ud x_1\,
\;
\int_{\frac{\ell_2-\pum}{N}}^{\frac{\ell_2+\pum}{N}}\ud x_2\; f(\bs{x})\;
\ ,
\label{AWSM_7}
\end{equation}
so that varying $f\in\Al$ yields $\mathrm{Ran}\pt{{\cal J}_{\nh,\infty}}={\cal
D}_\nh$. In order to recast~\eqref{AWSM_7} into a nicer expression, 
we introduce\\[-2ex]
\begin{quote}
\begin{DDD}[\co{R}unning \co{A}verage \co{O}perator (\co{RAO})]{}\ 
\label{runave}
\\[-0.5ex]
	Let $Q_N\pt{\bs{x}}$ denote the square of
	side $1/N$, oriented parallel to the axis
	of the torus and centered around $\bs{x}$;
	then, the \co{R}unning
	\co{A}verage \co{O}perator\\ $\Gamma_N :  \Lspace{\infty}{\cal
	X}\longmapsto \Cspace{0}{\IT}$, is defined by
\begin{equation}
\Lspace{\infty}{\IT}
\ni f(\bs{x})\longmapsto\Gamma_N\pt{f}(\bs{x})=:
N^2 
\;
\int_{Q_N\pt{\bs{x}}} \mu(\ud\bs{y})\;f(\bs{y})\;
\in\Cspace{0}{\IT}
\ \cdot\notag
\end{equation}
\end{DDD}
\newpage
\begin{PPP}{}\ 
\label{runaveprps}
\\[-1ex]
	Given $f\in \Lspace{\infty}{\IT}$, the function
$f_N^{(Q)}\coleq\Gamma_N\pt{f}$ is uniformly \mbox{continuous on $\IT$};
	moreover, the \co{R}unning \co{A}verage \co{O}perator has norm
\begin{equation}
\norm{\Gamma_N}{\cal B}\coleq\sup_{f\in \Lspace{\infty}{\IT}}
\frac{\norm{\Gamma_N\pt{f}}{0}}{\norm{f}{\infty}}=1\ \cdot
\label{AWSM_8}
\end{equation}
\end{PPP}
\end{quote}\ \\[-8ex]
\noindent
\textbf{Proof:}\\[3ex]
Let $\bs{x}_0\in\IT$, $\bs{x}\in Q_N\pt{\bs{x}_0}$ and $\car_E$ denote
the characteristic function of  
$E\subset\IT$. By Definition~\eqref{runave}:
\begin{align}
\abs{f_N^{(Q)}\pt{\bs{x}_0}-f_N^{(Q)}\pt{\bs{x}}}
&=N^2\;
\abs{\int_{\IT}\mu(\ud\bs{y})\;f(\bs{y})\;
\pt{\car_{Q_N\pt{\bs{x}_0}}(\bs{y})-{\cal
X}_{Q_N\pt{\bs{x}}}(\bs{y})} 
}\notag\\
&\leq N^2\;\norm{f}{\infty}
\int_{\IT}\mu(\ud\bs{y})\;
\abs{\car_{Q_N\pt{\bs{x}_0}}(\bs{y})-{\cal
X}_{Q_N\pt{\bs{x}}}(\bs{y})}\notag\\
&=N^2\;\norm{f}{\infty}\bigg[
\mu\Big(Q_N\pt{\bs{x}_0}\cup Q_N\pt{\bs{x}}\Big)\;-\;
\mu\Big(Q_N\pt{\bs{x}_0}\cap Q_N\pt{\bs{x}}\Big)
\bigg]\ \cdot\notag
\end{align}
According to our hypothesis, $\bs{x}\in
Q_N\pt{\bs{x}_0}$, thus geometrical considerations lead to:
\begin{align}
\mu\Big(Q_N\pt{\bs{x}_0}\cup Q_N\pt{\bs{x}}\Big)&\leq
\Big(\frac{1}{N}+\abs{x_1-x_{01}}\Big)
\Big(\frac{1}{N}+\abs{x_2-x_{02}}\Big)\notag\\
\mu\Big(Q_N\pt{\bs{x}_0}\cap Q_N\pt{\bs{x}}\Big)&=
\Big(\frac{1}{N}-\abs{x_1-x_{01}}\Big)
\Big(\frac{1}{N}-\abs{x_2-x_{02}}\Big)\notag\\
\mu\Big(Q_N\pt{\bs{x}_0}\cup Q_N\pt{\bs{x}}\Big)\;& -\;
\mu\Big(Q_N\pt{\bs{x}_0}\cap Q_N\pt{\bs{x}}\Big)\leq
\frac{2}{N}\Big(\abs{x_1-x_{01}}+\abs{x_2-x_{02}}\Big)\notag\\
& \phantom{-\;\ \,
\mu\Big(Q_N\pt{\bs{x}_0}\cap Q_N\pt{\bs{x}}\Big)}\leq\;
\frac{2\sqrt{2}}{N}\;
\norm{\bs{x}_0-\bs{x}}{}\ ,\notag
\end{align}
so that
$\displaystyle \abs{f_N^{(Q)}\pt{\bs{x}_0}-f_N^{(Q)}\pt{\bs{x}}}\leq
2\sqrt{2}\;N\;\norm{f}{\infty}\;\norm{\bs{x}_0-\bs{x}}{}$,
which proves the continuity of $f_N^{(Q)}$, while uniform continuity
comes from $\IT$ being compact.

Concerning the norm in~\eqref{AWSM_8}, 
the upper bound $\norm{\Gamma_N}{\cal B}\leq 1$ is clear and the
maximum is reached by choosing $f$ constant.\hfill$\qed$\\[3ex]

\noindent By means of the \co{RAO}, the discretization operator
in~\eqref{AWSM_1} can be conveniently written as  
\begin{equation}
{\cal J}_{\nh,\infty}(f)
=\sum_{\bs{\ell} \in {(\ZNZ{N})^2}} 
f_N^{(Q)}\pt{\frac{\bs{\ell}}{N}}
\ket{\bs{\ell}}\bra{\bs{\ell}}
\label{AWJNI1}
\ \cdot
\end{equation}
Analogously, 
the de--discretization operator
in~\eqref{AWSM_2} can be recast as  
\begin{equation}
{\cal J}_{\infty,\nh}(X)(\bs{x})
= 
\sum_{\bs{\ell} \in {(\ZNZ{N})^2}} 
X_{\bs{\ell},\bs{\ell}}\;
\;\delta^{(N)}_{\bs{\ell}\:,\:\hat{\bs{x}}_N}\;
=
\sum_{\bs{\ell} \in {(\ZNZ{N})^2}} 
X_{\bs{\ell},\bs{\ell}}\;{\cal
X}_{Q_N\pt{\frac{\bs{\ell}}{N}}}(\bs{x})
\label{AWJIN1due}
\ ,
\end{equation}
thus proving that $\mathrm{Ran}\pt{{\cal J}_{\infty,\nh}}={\cal S}\pt{\IT}$.

Moreover, combining equations~\eqref{AWJNI1} and ~\eqref{AWJIN1due},
we explicitly get the simple function arising from 
$f\in\Lspace{\infty}{\IT}$, via \co{AW}
discretization/de--discretization:
\begin{equation}
\pt{{\cal J}_{\infty,\nh}\circ{\cal J}_{\nh,\infty}}(f)(\bs{x})
=\sum_{\bs{\ell} \in {(\ZNZ{N})^2}} 
\Gamma_N\pt{f}\pt{\frac{\bs{\ell}}{N}}
\;\car_{Q_N\pt{\frac{\bs{\ell}}{N}}}(\bs{x})
\label{AWJINJNI}
\ \cdot
\end{equation}
The action of the operator ${\cal J}_{\infty,\nh}\circ{\cal
J}_{\nh,\infty}$ can be seen in Figures~\ref{due}
and~\ref{tre}.
\section{Proof of Proposition~\ref{Lemma2}}
\label{app_Lem_uno}
\vspace{3mm}
We start by proving the inclusions~\eqref{lemma_1_5}.

For every real number $t$, we have $0\leq\bk{N t
+\frac{1}{2}}=N t
+\frac{1}{2}-\floor{N t
+\frac{1}{2}}<1$, so that
\begin{alignat}{3}
\abs{t-\frac{\floor{Nt+\frac{1}{2}}}{N}}&\leq\frac{1}{2N}
\notag
 \ &\ \quad&,\quad\ &\ 
\forall\;t&\in\IR  \cdot
\intertext{From~\eqref{loc_c32} in Definition~\ref{xxxnnn}, we derive}
 d_{\IT}\pt{\bs{x}\;,\;\frac{\hat{\bs{x}}_N}{N}}&\leq\frac{1}{\sqrt{2}N}
\label{nuovopt_2}
\ &\ \quad&,\quad\ &\ 
\forall\;\bs{x}&\in\IT \cdot 
\end{alignat}
Then, let us consider the triangular inequality
\begin{align}
d_{\IT}\pt{\bs{x}\;,\;\bs{y}}&\leq
d_{\IT}\pt{\bs{x}\;,\;\frac{\hat{\bs{x}}_N}{N}} +
d_{\IT}\pt{\frac{\hat{\bs{x}}_N}{N}\;,\;\bs{y}}\qquad\forall\;\bs{y}\in\IT
\qquad 
\label{nuovopt_3}
\ ,
\intertext{and let us take the infimum over the set $\bs{y}\in\Gamma_n$ defined
in~\eqref{Gamman}}
d_{\IT}\pt{\frac{\hat{\bs{x}}_N}{N}\;,\;\Gamma_n}
&\geq  
d_{\IT}\pt{\bs{x}\;,\;\Gamma_n} -
d_{\IT}\pt{\bs{x}\;,\;\frac{\hat{\bs{x}}_N}{N}} 
\notag
\\
&\geq  
d_{\IT}\pt{\bs{x}\;,\;\Gamma_n}-
\frac{1}{\sqrt{2}N}
\notag
\ ,
\intertext{where we used~\eqref{nuovopt_2}. Therefore, considering the
complement $\pq{\overline{\Gamma}_n\pt{\varepsilon}}^{\circ}$ of
the union of strip of width $\varepsilon$, $\overline{\Gamma}_n\pt{\varepsilon}$
defined in~\eqref{Gnbar_1}, we get that}
\bs{x}\in\pq{\overline{\Gamma}_n\pt{\varepsilon}}^{\circ}\quad&\Longrightarrow\quad
\frac{\hat{\bs{x}}_N}{N}
\in\pq{\overline{\Gamma}_n\pt{\varepsilon-\frac{1}{\sqrt{2}N}}}^{\circ}\
\cdot
\notag
\end{align}
Further, from~\eqref{Gnbar_2}, it follows that, if the lattice point
$\frac{\hat{\bs{x}}_N}{N}$ 
does not belong to
$\overline{\Gamma}_n\pt{\varepsilon-\frac{1}{\sqrt{2}N}}$, then the
corresponding point $\bs{x}\in\IT$ must belong to
${G_n^N}\pt{\varepsilon-\frac{1}{\sqrt{2}N}}$.\\
Changing $\varepsilon-\frac{1}{\sqrt{2}N}\longmapsto\varepsilon$ we
obtain the first inclusion relation in equation~\eqref{lemma_1_5};
the second one follows by interchanging the
role played by $\frac{\hat{\bs{x}}_N}{N}$ and $\bs{x}$
in~\eqref{nuovopt_3}.

In order to prove~\eqref{lemma1_1} , we start by considering the matrices 
	$S_\alpha=\pt{\begin{smallmatrix} 
	1+ \alpha  & 1\\ \alpha & 1 \end{smallmatrix}}$ and its inverse
	$S_\alpha^{-1}=\pt{\begin{smallmatrix} 
	1 & - 1\\ -\alpha & 1 + \alpha \end{smallmatrix}}$. Let $\eta$
	be the
	largest (positive) eigenvalue of $\sqrt{S_\alpha^\dagger
S_\alpha^{\phantom{\dagger}}}$; its characteristic polynomial for $\eta$ is $\displaystyle
\eta^4 -\pt{2\:\alpha^2 + 2\:\alpha + 3}\eta^2 + 1 = 0$, whence
$\eta$ attains its minimum $\eta_{\text{min}}=\sqrt{2}$ at
$\alpha=-\frac{1}{2}$. Then, we set
$\widetilde{N}\coleq 2\,\sqrt{2}\pt{\sqrt{2}+1}\eta^{2n}$,
$n\in\IN$, choose $N>\widetilde{N}$ and proceed by induction.\\[2ex]
\noindent $\bs{p=0}:$ \quad from
definitions~\eqref{Ualpha} and~\eqref{Valpha}, it follows
\begin{equation}
d_{\IT}\pt{
\frac{U_\alpha^0\pt{N\bs{x}}}{N}\,,\,
\frac{V_\alpha^0\pt{\hat{\bs{x}}_N}}{N}}
 =d_{\IT}\pt{\bs{x}\,,\,\frac{\hat{\bs{x}}_N}{N}}<
\frac{1}{\sqrt{2}N}<
\frac{\sqrt{2}}{N}
\notag
\ ,
\end{equation}
where the first inequality follows from~\eqref{nuovopt_2}, 
thus relation~\eqref{lemma1_1} holds for $p=0$.\\[2ex]
$\bs{p=q-1\,,\ 1 \leq q\leq n}:$ \quad since
\begin{multline}
d_{\IT}\pt{\frac{U_\alpha^q\pt{N\bs{x}}}{N},
\frac{V_\alpha^q\pt{\hat{\bs{x}}_N}}{N}}\leq
d_{\IT}\pt{\frac{U_\alpha\pt{U_\alpha^{q-1}\pt{N\bs{x}}}}{N},
\frac{U_\alpha\pt{V_\alpha^{q-1}\pt{\hat{\bs{x}}_N}}}{N}}+\\
+d_{\IT}\pt{\frac{U_\alpha\pt{V_\alpha^{q-1}\pt{\hat{\bs{x}}_N}}}{N},
\frac{V_\alpha\pt{V_\alpha^{q-1}\pt{\hat{\bs{x}}_N}}}{N}}\ ,
\notag
\end{multline}
using~\eqref{Ualpha} in the first term and noting that, from
definitions~\eqref{Ualpha} and~\eqref{Valpha}, the second term is
less or equal to $\frac{\sqrt{2}}{N}$, we get
\begin{equation}
d_{\IT}\pt{\frac{U_\alpha^q\pt{N\bs{x}}}{N},
\frac{V_\alpha^q\pt{\hat{\bs{x}}_N}}{N}}\leq
d_{\IT}\pt{
S_\alpha\pt{\frac{U_\alpha^{q-1}\pt{N\bs{x}}}{N}},
S_\alpha\pt{\frac{V_\alpha^{q-1}\pt{\hat{\bs{x}}_N}}{N}}
}+
\frac{\sqrt{2}}{N}\ \cdot
\notag
\end{equation}
By the induction hypothesis we have:
\begin{align}
d_{\IT}\pt{
\frac{U_\alpha^{q-1}\pt{N\bs{x}}}{N},
\frac{V_\alpha^{q-1}\pt{\hat{\bs{x}}_N}}{N}}
& \leq\frac{\sqrt{2}}{N}\pt{\frac{\eta^{q}-1}{\eta-1}}
\label{prova5_7}
\\
& \leq\frac{\sqrt{2}}{N}\;
\frac{1}{\sqrt{2}-1}\;\eta^q
\notag
\\
\Big(\eta>\sqrt{2}\ ,\ 1\leq q\leq n\quad\Longrightarrow\Big)\quad\quad
&<\frac{1}{2}\;\eta^{q-2n} <\frac{1}{2}\eta^{-1}\ \cdot
\label{prova5_9}
\intertext{Now we set $\varepsilon=\frac{\widetilde{N}}{2N}$, taking
 into account that $\eta\geq\sqrt{2}$ and use the right inclusion
 in~\eqref{lemma_1_5} to deduce that} 
\bs{x}\in
{G_n^N}\pt{\frac{\widetilde{N}}{2N}}
& \Longrightarrow
\bs{x}\not\in\overline{\Gamma}_n\pt{\frac{\widetilde{N}}{2N}-
\frac{1}{\sqrt{2}N}}
\notag
\ .
\end{align}
At this point, we make use of the following result, which shall be
proved in Lemma~\ref{Lemma3}.3: it states that if a point does not
belong to $\overline{\Gamma}_n\pt{\varepsilon}$, the union of the the
strips of width 
$\varepsilon\leq\frac{1}{2}$ up to 
time $n$, then its orbit under $S_\alpha$ up to time $n-1$ is farther 
away than $\varepsilon\eta^{-q}, 0\leq q<n$ from the discontinuity
line $\gamma_0$. Explicitly
\begin{equation*}
\bs{x} \not\in \overline{\Gamma}_n\pt{\varepsilon} \Longrightarrow
d_{\IT}\pt{S_\alpha^q\pt{\bs{x}},\gamma_0}>
\varepsilon\;\eta^{-q}   
\ ,\ \forall \ 0\leq q < n \ ,
\end{equation*}
whence
\begin{align}
d_{\IT}\pt{\frac{U_\alpha^{q-1}\pt{N\bs{x}}}{N},\gamma_0} & >
\pt{\frac{\widetilde{N}}{2N}-
\frac{1}{\sqrt{2}N}}\eta^{1-q}
\notag
 \\
& >
\frac{\sqrt{2}}{N}\pt{\frac{\eta^{2n-1}-\eta^{q-1}}{\eta-1}}\eta^{1-q}
\notag
 \\
& \geq \frac{\sqrt{2}}{N}\pt{\frac{\eta^{q}-1}{\eta-1}}\ ,
\label{prova5_10c}
\end{align}
where the second inequality comes from $\eta\geq\sqrt{2}$,
the relation $\displaystyle
\frac{\eta^n-1}{\eta-1}\leq\frac{1}{\sqrt{2}-1}\:\eta^n$ and the
following estimates:
\begin{align}
\pt{\frac{\widetilde{N}}{2N}-
\frac{1}{\sqrt{2}N}}
& =\frac{\sqrt{2}}{N}\pt{\pt{\sqrt{2}+1}\eta^{2n}
-\frac{1}{2}}\nonumber\\
& \geq   
\frac{\sqrt{2}}{N}\pq{
\pt{\sqrt{2}+1}\pt{\sqrt{2}-1}
\frac{\eta^{2n}-\eta+\eta-1}{\eta-1}
-\frac{1}{2}}\nonumber\\
& \geq\frac{\sqrt{2}}{N}\pq{
\;\eta\;
\frac{\eta^{2n-1}-1}{\eta-1}
+\frac{1}{2}}\geq\frac{\sqrt{2}}{N}
\pt{\frac{\eta^{2n-1}-\eta^{q-1}}{\eta-1}}\ \nonumber\cdot
\end{align}

Therefore, comparing~\eqref{prova5_10c} with~\eqref{prova5_7}
\begin{align}
d_{\IT}\pt{
\frac{U_\alpha^{q-1}\pt{N\bs{x}}}{N},
\frac{V_\alpha^{q-1}\pt{\hat{\bs{x}}_N}}{N}}
& < d_{\IT}\pt{\frac{U_\alpha^{q-1}\pt{N\bs{x}}}{N},\gamma_0} \quad,\quad\forall q\leq
n\ \cdot
\notag
\end{align}
As a consequence, the segment $\pt{
\frac{U_\alpha^{q-1}\pt{N\bs{x}}}{N},
\frac{V_\alpha^{q-1}\pt{\hat{\bs{x}}_N}}{N}}$ cannot cross the line
$\gamma_0$.
This condition, together with~\eqref{prova5_9}, allows
us to use another result proved in Lemma~\ref{Lemma3}.1b, which states
that if a segment $(A,B)$ on the torus does not cross the
discontinuity line $\gamma_0$ then $d_{\IT}\pt{S_\alpha\pt{A},
	S_\alpha\pt{B}} \leq 
	\eta\;d_{\IT}\pt{A,B}$. We can finally conclude with:
\begin{equation}
d_{\IT}\pt{\frac{U_\alpha^q\pt{N\bs{x}}}{N},
\frac{V_\alpha^q\pt{\hat{\bs{x}}_N}}{N}}\leq
\eta\frac{\sqrt{2}}{N}\pt{\frac{\eta^{q}-1}{\eta-1}}
+
\frac{\sqrt{2}}{N} =
\frac{\sqrt{2}}{N}\pt{\frac{\eta^{q+1}-1}{\eta-1}}\ \cdot\hfill\tag*{\qed}
\end{equation}
The following Lemma, which has been used in the proof of the previous
Proposition, deals with the geometrical properties of the
\co{S}awtooth dynamics.
\begin{quote}\ \\[-7ex]
\begin{LLL}{}\ \\[-6ex]
\label{Lemma3}
\begin{Ventry}{3)}
	\item[] 
 	With $\eta$ the largest (positive) eigenvalue of
	$\sqrt{S_\alpha^{\dagger}S_\alpha}$ and $A,B\in\IT$ such
	that 
$d_{\IT}\pt{A,B}<\frac{1}{2}\;\eta^{-1}$, it follows:\\[2ex]
\begin{subequations}
	\noindent 
\label{t2r2tot}
	(1a) If the segment $(A,B)$ does not cross $\gamma_{-1}$, then 
	\begin{align}
	d_{\IT}\pt{S_\alpha^{-1}\pt{A},S_\alpha^{-1}\pt{B}} & \leq
	\eta\;d_{\IT}\pt{A,B}\ \cdot
\label{t2r2}
	\intertext{(1b) If $(A,B)$ does not cross
	$\gamma_{0}$, then} 
	d_{\IT}\pt{S_\alpha\pt{A},
	S_\alpha\pt{B}} & \leq 
	\eta\;d_{\IT}\pt{A,B}\ \cdot
\label{t2r2b}
	\end{align}
\end{subequations}
	\item[(2)] For any given $\alpha\in\IR$, $p\in{\IN}^+$ and 
	$0\leq\varepsilon\leq\frac{1}{2}\,\eta^{-1}$, 
\begin{equation}
\bs{x}\in\overline{\gamma}_{p-1}\pt{\varepsilon}
\Longrightarrow S_\alpha^{-1}\pt{\bs{x}}\in\pt{
\overline{\gamma}_p\pt{\eta\,\varepsilon}\cup
\overline{\gamma}_0\pt{\eta\,\varepsilon}}\ \cdot
\notag
\end{equation}
	\item[(3)] For any given $\alpha\in\IR$, $n\in{\IN}^+$ and
	$0\leq\varepsilon\leq\frac{1}{2}$, with $U_\alpha^q$ as
	in~\eqref{Ualpha},
\begin{equation}
\bs{x} \not\in \overline{\Gamma}_n\pt{\varepsilon} \Longrightarrow
d_{\IT}\pt{\frac{U_\alpha^q\pt{N\bs{x}}}{N},\gamma_0}>
\varepsilon\;\eta^{-q}   
\ ,\ \forall \ 0\leq q < n \ \cdot
\notag
\end{equation}
\end{Ventry}
\end{LLL}
\end{quote}
\noindent 
\textbf{Proof:}\\[3ex]
In the course of the proof, we shall use that
\begin{subequations}
\label{lemma1_3}
 
\begin{align}
\norm{S_\alpha^{\pm 1} \cdot \bs{v}}{{\IR}^2}&\leq
\eta^{\phantom{-1}}\norm{\bs{v}}{{\IR}^2}\ 
,
\label{lemma1_3a}
\\
\norm{S_\alpha^{\pm 1} \cdot \bs{v}}{{\IR}^2}&\geq
\eta^{-1}\norm{\bs{v}}{{\IR}^2}\ ,
\label{lemma1_3b}
\end{align}
\end{subequations}
which directly follows from the definition of $\eta$, where $\bs{v}$ is
any 2--dimensional real vector.

In order to prove~\eqref{t2r2tot}, it is
convenient to unfold $\IT$ and the discontinuity of $S_\alpha$ on the
plane $\IR^2$. This is most easily done as follows. Points
$A\in\IT={\IR}^2/{\IZ}^2$ are represented by equivalence classes 
\begin{equation}
\pq{\bs{a}}\coleq\pg{\bs{a}+\bs{n}\ ,\ \bs{n}\in{\IZ}^2}\ ,\
\bs{a}\in{[0,1)}^2 \ \cdot
\label{eqclass}
\end{equation}
\noindent Given $A,B\in\IT$, let $A^{b}\in\pq{\bs{a}}$ be such that
\begin{equation}
d_{\IT}\pt{\pq{\bs{a}},\pq{\bs{b}}} = \norm{A^b-\bs{b}}{{\IR}^2}\ \cdot
\notag
\end{equation}
Notice that
\begin{equation}d_{\IT}\pt{\pq{\bs{a}},\pq{\bs{b}}}
=\norm{\bs{a}-\bs{b}}{{\IR}^2} 
\quad 
\text{iff} \quad \norm{\bs{a}-\bs{b}}{{\IR}^2}\leq\frac{1}{2}
\label{normITIR}
\end{equation}
($1$a) $\pt{A,B}$ not crossing
$\gamma_{-1}$ means that the segment 
$\pt{A^b_{\phantom{2}},\bs{b}}$ does not intersect $\gamma_{-1}$.
Periodically covering the plane--${\IR}^2$ by squares ${[0,1)}^2$, 
the $\gamma_{-1}$-lines form a set of (parallel) straight lines $x_1 - x_2 =
n\in\IZ$; it follows that 
$\pt{A^b_{\phantom{2}},\bs{b}}$ does not cross
$\gamma_{-1}$ iff
\begin{equation}
\floor{A^b_1 - A^b_2} = \floor{b_1 - b_2}\ ,
\label{ldt2_2}
\end{equation}
where the integral part on the r.h.s. takes values $0,-1$,  depending
on which side of the diagonal $\gamma_{-1}$ the point $\bs{b}$ lies
within.\\
As $S_\alpha^{\pm}$ are not sensitive to the integer part of their
arguments, their actions are the same on all elements of the
equivalence classes~\eqref{eqclass}, that is
\begin{align}
d_{\IT} \pt{S_\alpha^{-1}\pt{A},S_\alpha^{-1}\pt{B}}  & = 
d_{\IT} \pt{S_\alpha^{-1}\pt{[\bs{a}]},S_\alpha^{-1}\pt{[\bs{b}]}}  = 
d_{\IT}\pt{S_\alpha^{-1}\pt{A^b_{\phantom{2}}},S_\alpha^{-1}\pt{\bs{b}}}\
\cdot 
\notag
\intertext{By expanding $\bk{x} = x - \floor{x}$, using the definition
of $S_\alpha^{-1}\pt{\cdot}$ and putting together all
integral contributions, condition~\eqref{ldt2_2} yields}
d_{\IT} \pt{S_\alpha^{-1}\pt{A},S_\alpha^{-1}\pt{B}}  & 
=\min_{\bs{m}\in{\IZ}^2}\norm{
S_\alpha^{-1}\pt{A} - S_\alpha^{-1}\pt{B}
+\bs{m}
}{{\IR}^2} \nonumber\\
& = \min_{\bs{m^{\prime}}\in{\IZ}^2}\norm{
S_\alpha^{-1}\cdot
\pt{A^b - \bs{b}}
+\bs{m^{\prime}}
}{{\IR}^2}
\notag\\
& = d_{\IT}\pt{S_\alpha^{-1}\cdot\pt{A^b_{\phantom{2}} - \bs{b}},0}\ \cdot
\notag
\end{align}
Applying~\eqref{lemma1_3a}, since we assumed
$d_{\IT}\pt{A,B}<\frac{1}{2}\;\eta^{-1}$, we estimate 
\begin{align}
\norm{S_\alpha^{-1}\cdot\pt{A^b_{\phantom{2}} - \bs{b}}}{{\IR}^2}
& \leq \eta \norm{A^b_{\phantom{2}} - \bs{b}}{{\IR}^2}
\notag
\\
= \eta\;d_{\IT}\pt{A,B} < \frac{1}{2}\ \cdot
\notag
\intertext{In particular, using~\eqref{normITIR},
the previous inequalities imply} 
d_{\IT}\pt{S_\alpha^{-1}\cdot\pt{A^b_{\phantom{2}} - \bs{b}},0}
& = \norm{S_\alpha^{-1}\cdot\pt{A^b_{\phantom{2}} - \bs{b}}}{{\IR}^2}\ 
\leq \eta\;d_{\IT}\pt{A,B} \ 
\cdot
\notag
\end{align}
\noindent ($1$b) Using the same argument as (1a), the union of
$\gamma_{0}$-lines constitute a set of 
straight lines $x_1= n\in\IZ$; 
Therefore the segment $\pt{A^b_{\phantom{2}},\bs{b}}$ does not cross
$\gamma_{0}$ iff
\begin{equation}
\floor{A^b_1} = \floor{b_1}\ \cdot
\label{ldt2_2p}
\end{equation}
As done before, by means of~\eqref{ldt2_2p}, we arrive at 
\begin{equation}
d_{\IT} \pt{S_\alpha\pt{A},S_\alpha\pt{B}}  = 
d_{\IT} \pt{S_\alpha\pt{A^b_{\phantom{2}}},S_\alpha\pt{\bs{b}}}  
= d_{\IT}\pt{S_\alpha\cdot\pt{A^b_{\phantom{2}} - \bs{b}},0}\
\cdot
\notag
\end{equation}
The proof can now be completed exactly as for point
($2$a) before.\\[2.5ex]
(2) We denote by $\displaystyle d_{\IT}\pt{\bs{x},\gamma} = 
\inf_{\bs{y}\in\gamma}
d_{\IT}\pt{\bs{x},\bs{y}}$ the distance of the point $\bs{x}\in
\IT$ from a curve $\gamma\in \IT$. Then, from
Definition~\eqref{Gnbar_0} we have: 
\begin{equation} 
\bs{x}\in\overline{\gamma}_{p-1}\pt{\varepsilon}
\Longrightarrow \varepsilon \geq
d_{\IT}\pt{\bs{x},\gamma_{p-1}} 
= d_{\IT}\pt{\bs{x},\bs{y}^{\star}}\ ,
\label{distg_1}
\end{equation} 
where $\bs{y}^{\star}$ 
is the nearest point to $\bs{x}$ belonging to $\gamma_{p-1}$.\\
We distinguish two cases:
\begin{quote}
\begin{Ventry}{($2^{\prime\prime}$)}
	\item[($2^{\prime}$)] \underline{The segment
	$\pt{\bs{x},\bs{y}^{\star}}$ does not
cross\protect\footnotemark \ $\gamma_{-1}$}\\[2ex] 
	From~\eqref{distg_1} and point ($1$a), since
        $S_\alpha^{-1}\pt{\bs{y}^{\star}}\in\gamma_p$
        (see~\eqref{gammakka}),
	we get
\begin{align} 
d_{\IT}\pt{S_\alpha^{-1}\pt{\bs{x}},\gamma_p}
& \leq
d_{\IT}\pt{S_\alpha^{-1}\pt{\bs{x}},
S_\alpha^{-1}\pt{\bs{y}^{\star}}}\notag\\
&\leq \eta\:d_{\IT}\pt{\bs{x},\bs{y}^{\star}}\leq
\eta\:\varepsilon\ \cdot
\notag
\end{align}
\protect\footnotetext{we stipulate that, if
$\bs{y}^{\star}\in\gamma_{-1}$ or $\bs{x}\in\gamma_{-1}$, we are still
in a non--crossing condition} 
Therefore
$\displaystyle
S_\alpha^{-1}\pt{\bs{x}}\in\overline{\gamma}_p\pt{\eta\,\varepsilon}$. 
	\item[($2^{\prime\prime}$)] \underline{The segment
	$\pt{\bs{x},\bs{y}^{\star}}$ crosses $\gamma_{-1}$}.\\
In this case, there exists $\bs{z}\in\gamma_{-1}$ such that 
\begin{gather} 
d_{\IT}\pt{\bs{x},\bs{y}^{\star}} =
d_{\IT}\pt{\bs{x},\bs{z}} + d_{\IT}\pt{\bs{z},\bs{y}^{\star}}\
\cdot 
\label{distg_4}
\intertext{Then, from~\eqref{distg_1} and~\eqref{distg_4},}
\varepsilon \geq d_{\IT}\pt{\bs{x},\bs{y}^{\star}} \geq
d_{\IT}\pt{\bs{x},\bs{z}}\ \cdot
\notag
\end{gather}
Since, according to~\eqref{gammakk},
$S_\alpha^{-1}\pt{\bs{z}}\in\gamma_0$, from point ($1$a) we get 
\begin{equation} 
d_{\IT}\pt{S_\alpha^{-1}\pt{\bs{x}},\gamma_0}\leq
d_{\IT}\pt{S_\alpha^{-1}\pt{\bs{x}},S_\alpha^{-1}\pt{\bs{z}}}
\leq \eta\:\varepsilon\ ,
\notag
\end{equation}
that is $\displaystyle
S_\alpha^{-1}\pt{\bs{x}}\in\overline{\gamma}_0\pt{\eta\,\varepsilon}$.
\end{Ventry}
\end{quote}
(3) From point ($2$), it follows that, when
$0\leq\varepsilon\leq\frac{1}{2}$, for $p\in{\IN}^+$,
\begin{equation}
\bs{x}\not\in\pt{
\overline{\gamma}_p\pt{\varepsilon}\cup
\overline{\gamma}_0\pt{\varepsilon}}
\Longrightarrow
S_\alpha\pt{\bs{x}}\not\in\overline{\gamma}_{p-1}\pt{\eta^{-1}\varepsilon}\
\cdot 
\label{ddg0_1}
\end{equation}
We prove by induction that, when
$0\leq\varepsilon\leq\frac{1}{2}$, for $m\in{\IN}^+$,
\begin{equation}
\bs{x}\not\in\bigcup_{p=0}^m\;
\overline{\gamma}_p\pt{\varepsilon}
\Longrightarrow
S_\alpha\pt{\bs{x}}\not\in
\bigcup_{p=0}^{m-1}\;
\overline{\gamma}_p
\pt{\eta^{-1}\varepsilon}\ \cdot
\label{ddg0_2}
\end{equation}
For $m=1$,~\eqref{ddg0_2} follows from~\eqref{ddg0_1}; 
if~\eqref{ddg0_2} holds for $m=r$, then take
\begin{equation}
\bs{x}\not\in\bigcup_{p=0}^{r+1}\;
\overline{\gamma}_p\pt{\varepsilon}
\text{\ . This means that}\quad
\bs{x}\not\in\bigcup_{p=0}^{r}\;
\overline{\gamma}_p\pt{\varepsilon}\quad\text{and}\quad
\bs{x}\not\in\pt{\overline{\gamma}_{r+1}\pt{\varepsilon}\cup
\overline{\gamma}_0\pt{\varepsilon}}\ \cdot\nonumber
\notag
\end{equation}
Now, using the induction hypothesis and~\eqref{ddg0_1}, we get
\begin{equation}
\bs{x}\not\in\bigcup_{p=0}^{r+1}\;
\overline{\gamma}_p\pt{\varepsilon}
\Longrightarrow
S_\alpha\pt{\bs{x}}\not\in
\bigcup_{p=0}^{r-1}\;
\overline{\gamma}_p\pt{\eta^{-1}\varepsilon}
\quad\text{and}\quad
S_\alpha\pt{\bs{x}}\not\in
\overline{\gamma}_r\pt{\eta^{-1}\varepsilon}\ \cdot
\notag
\end{equation}
\noindent Setting $m=n-1$ and iterating $q$ times the
implication~\eqref{ddg0_2} argument, we get
\begin{align}
\bs{x}\not\in\bigcup_{p=0}^{n-1}\;
\overline{\gamma}_p\pt{\varepsilon}
& \Longrightarrow
S_\alpha^q\pt{\bs{x}}\not\in
\bigcup_{p=0}^{n-1-q}\;
\overline{\gamma}_p
\pt{\eta^{-q}\varepsilon}\quad ,\quad \forall \ 0\leq q < n\
\cdot
\notag
\intertext{In particular $\displaystyle S_\alpha^q\pt{\bs{x}}
\not\in\overline{\gamma}_0\pt{\eta^{-q}\varepsilon}$, which leads to
the lower bound} 
& d_{\IT}\pt{S_\alpha^q\pt{\bs{x}},\gamma_0}
>\eta^{-q}\varepsilon\quad ,\quad \forall \ 0\leq q < n \ ,
\notag
\end{align}
whence the result follows in view of Definitions~\eqref{Ualpha}
and~\eqref{Gnbar_1}.\hfill$\qed$
\section{Proof of Proposition~\ref{Lemma1}}
\label{app_Lem_due}
\vspace{3mm}
(a) In~\eqref{gammakka}, we have defined $\gamma_p =
S_\alpha^{-p}\pt{\gamma_0}$ where  
$S_\alpha^{-1}\pt{\bs{x}}$ (as well as $S_\alpha^{-p}\pt{\bs{x}}$)
is a piecewise continuous mapping onto $\IT$ with jump--discontinuities 
across the $\gamma_p$ lines
due to the presence of
the function $\bk{\cdot}$ in~\eqref{AoDC_1d}. Away from the 
discontinuities, $S_\alpha^{-p}\pt{\bs{x}}$ behaves as the matrix action 
$S_\alpha^{-p}\cdot\bs{x}$. We want now to estimate the length
$l\pt{\gamma_p}$; in order to do that, we unfold $\gamma_p$ on the
plane and calculate the length of the segment 
$\pg{\bs{x}\in{\IR}^2\ \Big|\ \bs{x}=S_\alpha^{-p}\cdot\pt{
\begin{smallmatrix}
0\\y
\end{smallmatrix}}
\ ,\ y\in [0,1)}$, which, in its turn, is 
the image of $\gamma_0$ under the matrix action given by
$S_\alpha^{-p}\cdot\bs{x}$. Therefore, using~\eqref{lemma1_3a}, the
result follows.\\[2.5ex]
(b) Let $\overline{L}\pt{\varepsilon}$ denote the set of points having
distance from 
a segment of length $L$ smaller
or equal than $\varepsilon$: it has a volume (under the Lebesgue
measure $\mu$) given by
\begin{equation*}
\mu\pt{\overline{L}\pt{\varepsilon}} = 2\,L\,\varepsilon +
\pi\varepsilon^2 \ , 
\end{equation*}
where the last term on the r.h.s. takes into account rounding of the
extremes of the strip by to semi--circle of radius $\varepsilon$.
Then~\eqref{lemma_1_2} follows from~\eqref{lemma_1_1}.\\[2.5ex]
(c) This follows from Definition~\eqref{Gnbar_1}:
\begin{align}
\mu\pt{\overline{\Gamma}_n\pt{\varepsilon}} & = 
\mu\pt{\bigcup_{p=0}^{n-1}\;
\overline{\gamma}_p\pt{\varepsilon}}
\leq \sum_{p=0}^{n-1}\;\mu\pt{
\overline{\gamma}_p\pt{\varepsilon}}\ \cdot\nonumber
\intertext{Using~\eqref{lemma_1_2}, we can write:}
\mu\pt{\overline{\Gamma}_n\pt{\varepsilon}} 
&\leq 2\,\varepsilon\sum_{p=0}^{n-1}\eta^p + 
\sum_{p=0}^{n-1}\pi\,\varepsilon^2 
 = 2\,\varepsilon \,\frac{\eta^n-1}{\eta^{\phantom{p}}-1}
+ n\,\pi\,\varepsilon^2\ \cdot\nonumber
\intertext{Finally, the estimate
$\displaystyle \frac{x^p-1}{x^{\phantom{p}}-1}\leq
\pt{\sqrt{2}+1}x^p$, valid for $x>\sqrt{2}$, yields}
\mu\pt{\overline{\Gamma}_n\pt{\varepsilon}} 
&\leq 2\,\varepsilon \pt{\sqrt{2}+1} \eta^n +
n\,\pi\,\varepsilon^2
&&
\ \cdot\nonumber
\end{align}\\[2.5ex]
(d) By writing the left inclusion in~\eqref{lemma_1_5} in terms of
complementary sets, with 
$\varepsilon=\frac{\widetilde{N}}{2N}$, we get:\\[-5.5ex]
\begin{align}
\pq{{G_n^N}\pt{\frac{\widetilde{N}}{2N}}}^{\circ}&\subseteq 
\overline{\Gamma}_n\pt{\frac{\widetilde{N}}{2N}+ 
\frac{1}{\sqrt{2}N}}\ \text{and so}
\notag
\\
\mu\pt{\pq{{G_n^N}\pt{\frac{\widetilde{N}}{2N}}}^{\circ}}&\leq
\mu\pt{\overline{\Gamma}_n\pt{\frac{\widetilde{N}}{2N}+ 
\frac{1}{\sqrt{2}N}}}
\notag
\ \cdot
\intertext{By substituting in~\eqref{lemma_1_3} 
$\frac{\widetilde{N}+\sqrt{2}}{2N}=\frac{\widetilde{N}}{2N}+  
\frac{1}{\sqrt{2}N}$ in the place of $\varepsilon$, we get:}
\mu\pt{\pq{{G_n^N}\pt{\frac{\widetilde{N}}{2N}}}^{\circ}}\leq
\frac{\widetilde{N}+\sqrt{2}}{2N}
& \pt{\sqrt{2}+1}
\pt{2\,\eta^n + \frac{n}{\sqrt{2}+1}\,\pi\,
\frac{\widetilde{N}+\sqrt{2}}{2N}
}\ \cdot
\label{nuovopt_9}
\end{align}
Finally, the r.h.s of~\eqref{nuovopt_9}, can be estimated by the
following upper bounds:
\begin{align}
\pi\,
\frac{\widetilde{N}+\sqrt{2}}{2N}& <2\nonumber
\,\\ 
\frac{n}{\sqrt{2}+1}
& <\eta^n\nonumber
\,\\ 
\pt{\widetilde{N}+\sqrt{2}}
\pt{\sqrt{2}+1}
& <19\;\eta^{2n}
\nonumber
\end{align}
which hold for $\displaystyle \forall\;N>\widetilde{N},\ 
\eta\geq\sqrt{2}$ and $\displaystyle \forall\; n\in\IN^+$.
This ends the proof.\hfill$\qed$
\singlespacing   		

\end{document}